\documentclass[10pt,twocolumn,twoside,journal]{IEEEtran}
\usepackage{subfigure}
\usepackage{setspace}
\usepackage{amsmath}
\usepackage{amssymb}
\usepackage{amsfonts}
\usepackage{amscd}
\usepackage{mathrsfs}
\usepackage[final]{graphicx}
\usepackage{graphicx}
\usepackage{psfrag}
\usepackage{url}
\usepackage{textcomp}
\usepackage{multirow}

\newcommand{\Define}{\stackrel{\triangle}{=}}

\begin{document}


\title{{\huge Low-Complexity Detection/Equalization in Large-Dimension 
MIMO-ISI Channels Using Graphical Models}
}
\author{{\large Pritam Som, Tanumay Datta, N. Srinidhi, A. Chockalingam, 
and B. Sundar Rajan}  \\
Department of ECE, Indian Institute of Science, Bangalore-560012, India.
}

\markboth{Pritam Som \MakeLowercase{\textit{et al.}}:
Low-Complexity Detection/Equalization in Large-Dimension MIMO-ISI Channels
Using Graphical Models
}
{Pritam Som \MakeLowercase{\textit{et al.}}: 
Low-Complexity Detection in Large-Dimension MIMO-ISI Channels
Using Graphical Models
}

\maketitle
\begin{abstract}
In this paper, we deal with low-complexity near-optimal detection/equalization 
in large-dimension multiple-input multiple-output inter-symbol interference 
(MIMO-ISI) channels using message passing on graphical models. A key 
contribution in the paper is the demonstration that near-optimal performance 
in MIMO-ISI channels with large dimensions can be achieved at low complexities 
through simple yet effective simplifications/approximations, although the 
graphical models that represent MIMO-ISI channels are fully/densely connected 
(loopy graphs). These include 1) use of Markov Random Field (MRF) based 
graphical model with pairwise interaction, in conjunction with {\em 
message/belief damping}, and 2) use of Factor 
Graph (FG) based graphical model with {\em Gaussian approximation of 
interference} (GAI). The per-symbol complexities are $O(K^2n_t^2)$ and 
$O(Kn_t)$ for the MRF and the FG with GAI approaches, respectively, 
where $K$ and $n_t$ denote the number of channel uses per frame, and number 
of transmit antennas, respectively. These low-complexities are quite 
attractive for large dimensions, i.e., for large $Kn_t$. From a performance 
perspective, these algorithms are even more interesting in large-dimensions 
since they achieve increasingly closer to optimum detection performance for 
increasing $Kn_t$. Also, we show that these message passing algorithms can 
be used in an iterative manner with local neighborhood search algorithms to 
improve the reliability/performance of $M$-QAM symbol detection. 
\end{abstract}


\section{Introduction}
\label{sec1}
Signaling in large dimensions can offer attractive benefits in wireless
communications. For example, transmission of signals using large spatial 
dimensions in multiple-input multiple-output (MIMO) systems with large 
number of transmit/receive antennas can offer increased spectral 
efficiencies \cite{fosc98}-\cite{paulraj}. The spectral efficiency in 
a V-BLAST MIMO system is $n_t$ symbols per channel use, where $n_t$ is 
the number of transmit antennas \cite{paulraj}. Severely delay-spread 
inter-symbol interference (ISI) channels can offer opportunities to harness 
rich diversity benefits \cite{proakis}. In an $L$-length ISI channel, 
each symbol in a frame is interfered by its previous $L-1$ symbols. 
However, the availability of $L$ copies of the transmitted signal in
ISI channels can be exploited to achieve $L$th order diversity. A
way to achieve this diversity is to organize data into frames, where
each frame consists of $K$ channel uses (i.e., $K$ dimensions in time), 
$K>L$, and carry out joint detection/equalization over the entire frame 
at the receiver. A MIMO-ISI channel with large $Kn_t$ and $L$ (referred 
to as large-dimension MIMO-ISI channel) is of interest because of
its potential to offer high spectral efficiencies (in large $n_t$) and 
diversity orders (in large $L$\footnote{A practical example of severely
delay-spread ISI channel with large $L$ is an ultra wideband (UWB)
channel \cite{ngoc}. UWB channels are highly frequency-selective,
and are characterized by severe ISI due to large delay spreads 
\cite{uwb0}-\cite{uwb3}. The number of multipath components (MPC) in 
such channels in indoor/industrial environments has been observed to 
be of the order of several tens to hundreds; number of MPCs ranging from 
12 to 120 are common in UWB channel models \cite{uwb0},\cite{uwb3}.}).
A major challenge, however, is detection complexity. The complexity 
of optimum detection is exponential in number of dimensions, 
which is prohibitive for large number of dimensions.
Our focus in this paper is to achieve near-optimal 
detection performance in large dimensions at low complexities. A 
powerful approach to realize this goal, which we investigate in this 
paper, is message passing on graphical models. 

Graphical models are graphs that indicate inter-dependencies between
random variables \cite{frey}. Well known graphical models include
Bayesian belief networks, factor graphs, and Markov random fields 
\cite{merl}. Belief propagation (BP) is a technique that solves 
inference problems using graphical models \cite{merl}. BP is a 
simple, yet highly effective, technique that has been successfully 
employed in a variety of applications including computational biology, 
statistical signal/image processing, data mining, etc. BP is well 
suited in several communication problems as well \cite{frey}; e.g., 
decoding of turbo codes and LDPC codes \cite{bp_turbo},\cite{ldpc}, 
multiuser detection in CDMA \cite{bpmud0}-\cite{bpmud2}, and MIMO 
detection \cite{ieee06}-\cite{itw10}.

Turbo equalization which performs detection/equalization and decoding 
in an iterative manner in coded data transmission over ISI channels have 
been widely studied \cite{douil_95},\cite{turbo_eq},\cite{teq_mag}. More 
recently, message passing on factor graphs based graphical models 
\cite{fg_sp} have been studied for detection/equalization on ISI 
channels \cite{euro_04}-\cite{wo}. 
In \cite{isi2}, it has been shown through simulations that application 
of sum-product (SP) algorithm to factor graphs in ISI channels converges 
to a good approximation of the exact a posteriori probability (APP) of the 
transmitted symbols. In \cite{fg_eq}, the problem of finding the linear 
minimum mean square error (LMMSE) estimate of the transmitted symbol 
sequence is addressed employing a factor graph framework. Equalization 
in MIMO-ISI channels using factor graphs are investigated in 
\cite{mimo_isi},\cite{wo}. In \cite{mimo_isi}, variable nodes of the 
factor graph correspond to the transmitted symbols, and each channel use
corresponds to a function node. Since the received signal at any channel
use depends on the past $L$ symbols transmitted from every transmit antenna, 
every function node is connected to $Ln_t$ variable nodes.  Near-MAP (maximum 
a posteriori probability) performance was shown through simulations for 
$n_t=2$ systems. However, the complexities involved in the computation of 
messages at the variable and function nodes are exponential in $Ln_t$, which 
are prohibitive for large spatial dimensions and delay spreads. In \cite{wo}, 
a Gaussian approximation of interference is used which significantly reduced 
the complexity to scale well for large $L$.
However, in terms of performance, the algorithm in \cite{wo} 
exhibited high error floors\footnote{Figure \ref{fig13a} shows an error 
floor in the approach in \cite{wo}. Whereas, in the same figure, our 
FG approach in Sec. \ref{sec4} is seen to avoid flooring and perform 
significantly better.}.

Our key contribution in this paper is the demonstration that graphical
models can be effectively used to achieve {\em near-optimal} 
detection/equalization
performance in {\em large-dimension} MIMO-ISI channels at {\em low 
complexities}. The achieved performance is good because 
detection is performed jointly over the entire frame of 
data; i.e., over the full $Kn_t\times 1$ data vector. While simple 
approximations/simplifications resulted in low 
complexities, the {\em large-dimension behavior}\footnote{We say that 
an algorithm exhibits `large-dimension behavior' if its bit error 
performance improves with increasing number of dimensions. The fact that 
turbo codes with BP decoding achieve near-capacity performance only when 
the {\em frame sizes are large} is an instance of large-dimension behavior.} 
natural in message passing algorithms contributed to the near-optimal 
performance in large dimensions. The graphical models we consider in this 
paper are Markov random fields (MRF) and factor graphs (FG). We show that 
these graphical models based algorithms perform increasingly closer to the 
optimum performance for increasing $n_t$ and increasing values of $K$ and 
$L$, keeping $L/K$ fixed. 

In the case of MRF approach (Section \ref{sec3}), we show that the use of 
{\em damping} of messages/beliefs, where messages/beliefs are computed as 
a weighted average of the message/belief in the previous iteration and the 
current iteration (details and associated references given in Section 
\ref{sec_3d}), is instrumental in achieving good performance. Simulation 
results show that the MRF approach exhibits large-dimension behavior, and 
that damping significantly improves the bit error performance (details given 
in Section \ref{sec_3f}). For 
example, the MRF based algorithm with message damping achieves close to 
unfaded single-input single-output (SISO) AWGN performance (which is a 
lower bound on the optimum detector performance) within 0.25 dB at 
$10^{-3}$ bit error rate (BER) in a MIMO-ISI channel with $n_t=n_r=4$, 
$K=100$ channel uses per frame (i.e., problem size is $Kn_t=400$ dimensions), 
and $L=20$ equal-energy multipath components (MPC). Similar performances 
are shown for large-MIMO systems with $n_t=n_r=16,32$ and $K=64$ (problem 
size $Kn_t=1024$ and 2048 dimensions). The per-symbol complexity of the 
MRF approach is $O(K^2n_t^2)$ (details in Section \ref{sec_3e}). 

In the case of FG approach (Section \ref{sec4}), the Gaussian 
approximation of interference (GAI) we adopt is found to be effective 
to further reduce the complexity by an order (Section \ref{sec_4a}); i.e., 
the per-symbol complexity of the FG with GAI approach is just $O(Kn_t)$, 
which is one order less than that of the MRF approach. The 
proposed FG with GAI approach is also shown to exhibit large-dimension 
behavior; its BER performance is almost the same as that of the 
MRF approach, and is significantly better than that of the scheme in 
\cite{wo} (Section \ref{sec_4b}). We also show that the proposed FG with 
GAI algorithm can be used in an iterative manner with local neighborhood 
search algorithms, like the reactive tabu search (RTS) algorithm 
in \cite{isi_gcom09}, to improve the performance of $M$-QAM detection 
(Section \ref{sec5}). 

Though the proposed algorithms are presented in the context of uncoded
systems, they can be extended to coded systems as well, through turbo
equalization \cite{douil_95}-\cite{teq_mag} \big(Receiver C in Fig. 1 of
\cite{teq_mag}\big) or through joint processing of the entire coded frame
using low-complexity graphical models \big(low-complexity approximations 
of Receiver A in Fig. 1 of \cite{teq_mag}\big). In \cite{bp_isit09}, we 
have investigated a scheme with separate MRF based detection followed by 
decoding \big(Receiver B is Fig. 1 of \cite{teq_mag}\big) in a $24\times 24$ 
large-MIMO system, and showed that a coded BER performance close to within
2.5 dB of the theoretical ergodic MIMO capacity is achieved. MIMO space-time 
coding schemes that can achieve separability of detection and decoding 
without loss of optimality \cite{blld} are interesting because they avoid 
the need for joint processing for optimal detection and decoding. If such 
detection-decoding separable space-time codes become available for 
large dimensions, the proposed algorithms can be applicable in their 
detection/equalization.   

The rest of the paper is organized as follows. In Section \ref{sec2}, we
present the considered MIMO system model in frequency selective fading. 
In Section \ref{sec3}, we present the proposed MRF based BP detector 
with damping and its BER performance in large dimensions. Section 
\ref{sec4} presents the FG with GAI based BP detector and its BER 
performance. In Section \ref{sec5}, the proposed hybrid RTS-BP algorithm 
for detection of $M$-QAM signals and its performance are presented. 
Conclusions are presented in Section \ref{sec6}.

\section{System Model}
\label{sec2}
We consider MIMO systems with cyclic prefixed single-carrier (CPSC) 
signaling, where the 
overall MIMO channel includes an FFT operation so that the transmitted 
symbols are estimated from the received frequency-domain signal (also 
referred to as SC-FDE: single-carrier modulation with frequency-domain 
equalization) \cite{cpsc1}-\cite{cpsc3}. Unlike OFDM signaling, CPSC 
signaling does not suffer from the peak to average power ratio 
(PAPR) problem. Also, CPSC with FD-MMSE equalizer performs better than 
OFDM at large frame sizes (large $K$) \cite{cpsc3}. We will see that 
our proposed BP based algorithms scale well for large dimensions in
MIMO-CPSC schemes (large $Kn_t$) and perform significantly better 
than MIMO-CPSC with FD-MMSE equalizer as well as MIMO-OFDM with MMSE/ML
equalizer. 

\begin{figure}
\includegraphics[width=3.45in, height=1.85in]{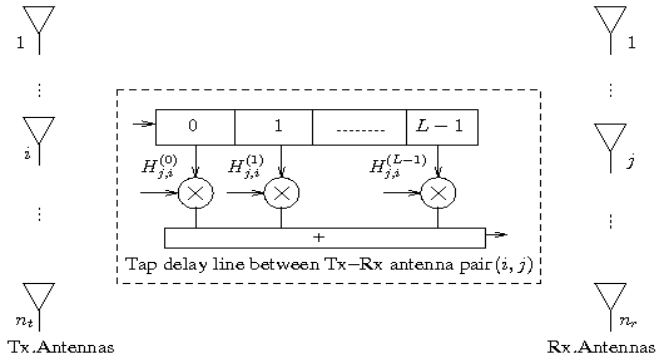}
\caption{MIMO-ISI Channel Model.}
\label{fig1}
\end{figure}

Consider a frequency-selective MIMO channel with $n_t$ transmit and
$n_r$ receive antennas as shown in Fig. \ref{fig1}. Let $L$ denote
the number of multipath components (MPC). Data is transmitted in 
frames, where each frame has $K'$ channel uses, out of which data
symbol vectors are sent in $K$ channel uses  $K\geq L$. These $K$
channel uses are preceded by a cyclic prefix (CP) of length $L-1$ 
channel uses so that $K'=K+L-1$. In each channel use, an $n_t$-length 
data symbol vector is transmitted using spatial multiplexing on $n_t$ 
transmit antennas. Let ${\bf x}_q \in \{\pm 1\}^{n_t}$ denote the 
data symbol vector transmitted in the $q$th channel use, $q=0,1,\cdots,K-1$. 
Though the symbol alphabet used here is BPSK, extensions to higher-order
alphabet are possible, and some are discussed later in the paper.
While CP avoids inter-frame interference, there will be ISI within the 
frame. The received signal vector at time $q$ can be written as
\begin{eqnarray}
\label{eq1}
{\mathbf y}_q &=& \sum_{l=0}^{L-1}{\mathbf H}_l\,{\mathbf x}_{q-l}+{{\mathbf w}}_q, \quad \,\, q=0,\cdots,K-1,
\end{eqnarray}
where ${\mathbf y}_q \in \mathbb{C}^{n_r}$,
${\mathbf H}_l \in \mathbb{C}^{n_r \times n_t}$ is the channel gain
matrix for the $l$th MPC such that $H_{j,i}^{(l)}$ denotes the entry
on the $j$th row and $i$th column of the ${\mathbf H}_l$ matrix, i.e.,
$H_{j,i}^{(l)}$ is the channel from $i$th transmit antenna to the $j$th 
receive antenna on the $l$th MPC. The entries of ${\mathbf H}_l$ are 
assumed to be i.i.d $\mathbb{C}\mathcal{N}(0,1)$. It is further assumed that 
${\mathbf H}_l$, $l=0,\cdots,L-1$ remain constant for one frame duration, 
and vary i.i.d from one frame to the other. 
${\mathbf w}_q \in \mathbb{C}^{n_r}$ is the additive white Gaussian noise 
vector at time $q$, whose entries are independent, each with variance 
$\sigma^2=n_tLE_s/\gamma$, where $\gamma$ is the average received SNR per
received antenna. The CP will render the linearly convolving channel to a 
circularly convolving one, and so the channel will be multiplicative in 
frequency domain. Because of the CP, the received signal in frequency 
domain, for the $i$th frequency index ($0 \leq i \leq K-1$), can be 
written as
\begin{eqnarray}
\label{eq3}
{\mathbf r}_i & = & {\mathbf G}_i \ {\mathbf u}_i + {\mathbf v}_i,
\end{eqnarray}
where
${\mathbf r}_i = \frac{1}{\sqrt{K}} \sum\limits_{q=0}^{K-1} e^{\frac{-2\pi {\mathbf j}qi}{K}} {\mathbf y}_q, \, \,$
{\normalsize ${\mathbf u}_i = \frac{1}{\sqrt{K}} \sum\limits_{q=0}^{K-1} e^{\frac{-2\pi {\mathbf j}qi}{K}} {\mathbf x}_q, \, \,$} 
${\mathbf v}_i = \frac{1}{\sqrt{K}} \sum\limits_{q=0}^{K-1} e^{\frac{-2\pi {\mathbf j}qi}{K}} {\mathbf w}_q, \, \,$
${\mathbf G}_i=\sum\limits_{l=0}^{L-1}e^{\frac{-2\pi{\mathbf j}li}{K}} {\mathbf H}_l$, and ${\mathbf j}=\sqrt{-1}$.
Stacking the $K$ vectors ${\mathbf r}_i$, $i=0,\cdots,K-1$, we write
\begin{eqnarray}
\label{eqChannelModel}
{\mathbf r} & = & \underbrace{{\mathbf{GF}}}_{\Define \,\, {\mathbf H}_{eff}}{\mathbf x}_{eff} + {\mathbf v}_{eff},
\end{eqnarray}
where
\[{\mathbf r}=\left[ \begin{array}{c} {\mathbf r}_0 \\ {\mathbf r}_1\\
\vdots \\ {\mathbf r}_{{\small K-1}} \end{array} \hspace{-2mm}\right]
\mathrm{,} \quad
{\mathbf G}=\left[ \begin{array}{cc} \begin{array}{ll} {\mathbf G}_0  & \\ & {\mathbf G}_1 \end{array} & {\mathbf 0} \\
{\mathbf 0}  &  \begin{array}{ll} \ddots & \\  & {\mathbf G}_{{\small K-1}} \end{array} \end{array}\hspace{-3mm} \right]
\mathrm{,} \quad \]
\[
{\mathbf x}_{eff}=\left[ \begin{array}{c} {\mathbf x}_0 \\ {\mathbf x}_1 \\ \vdots \\ {\mathbf x}_{{\small K-1}} \end{array} \hspace{-2mm}\right]
\mathrm{,} \quad
{\mathbf v}_{eff}=\left[ \begin{array}{c}{\mathbf v}_0 \\ {\mathbf v}_1\\ \vdots \\ {\mathbf v}_{{\small K-1}} \end{array}\right],
\]
\begin{eqnarray*}
{\mathbf F} & \hspace{-2mm} = & \hspace{-2mm}\frac{1}{\sqrt{K}}\left[ \begin{array}{llll} \rho_{{\small 0,0}}{\mathbf I}_{n_t} & \rho_{{\small 1,0}}{\mathbf I}_{n_t} & \cdots & \rho_{{\small K-1,0}}{\mathbf I}_{n_t} \\ \rho_{{\small 0,1}}{\mathbf I}_{n_t} & \rho_{{\small 1,1}}{\mathbf I}_{n_t} & \cdots & \rho_{{\small K-1,1}}{\mathbf I}_{n_t} \\
\vdots  & \vdots  & \cdots & \vdots   \\
\rho_{{\small 0,K-1}}{\mathbf I}_{n_t} & \rho_{{\small 1,K-1}}{\mathbf I}_{n_t} & \cdots & \rho_{{\small K-1,K-1}}{\mathbf I}_{n_t} \end{array} \hspace{-2mm}\right] \\
& = & \frac{1}{\sqrt{K}}{\mathbf D}_K \otimes {\mathbf I}_{n_t},
\end{eqnarray*}
where $\rho_{q,i}=e^{\frac{-2\pi{\mathbf j}qi}{K}}$, ${\mathbf D}_K$ is 
the $K$-point DFT matrix and $\otimes$ denotes the Kronecker product.
Equation (\ref{eqChannelModel}) can be written in an equivalent linear 
vector channel model of the form
\begin{eqnarray}
{\mathbf r} & = & {\mathbf H} {\mathbf x} + {\mathbf v},
\label{eqn1}
\end{eqnarray}
where ${\bf H}={\bf H}_{eff}$, ${\bf x}={\bf x}_{eff}$, and
${\bf v}={\mathbf v}_{eff}$. 
Note that the well known MIMO system model for flat fading can be 
obtained as a special case in the above system model with $L=K=1$.

We further note that, in the considered system, signaling is done along $K$ 
dimensions in time and $n_t$ dimensions in space, so that the total number 
of dimensions involved is $Kn_t$. We are interested in low-complexity 
detection/equalization in large dimensions (i.e., for large $Kn_t$) using 
graphical models. The goal is to obtain an estimate of vector ${\bf x}$, 
given ${\bf r}$ and the knowledge of ${\bf H}$. The optimal maximum 
a posteriori probability (MAP) detector takes the joint posterior 
distribution
\begin{eqnarray}
p({\bf x}\mid {\bf r},{\bf H}) & \propto & p({\bf r}\mid {\bf x},{\bf H}) \, p({\bf x}),\label{eqnd1}
\end{eqnarray}
and marginalizes out each variable as 
$p(x_i|{\bf r},{\bf H}) = \sum\limits_{x_{-i}} p({\bf x}|{\bf r},{\bf H})$,
where $x_{-i}$ stands for all entries of ${\bf x}$ except $x_i$.
The MAP estimate of the bit $x_i$, $i=1,\cdots,Kn_t$, is then given by
\begin{eqnarray}
{\widehat x}_i & = & {\mbox{arg max}\atop{a \in \{\pm 1\}}}
\,\, p \big(x_i = a \mid {\bf r},{\bf H}\big),
\label{MAPdetection}
\end{eqnarray}
whose complexity is exponential in $Kn_t$. In the following sections, we
present low-complexity detection algorithms based on graphical models
suited for the system model in (\ref{eqn1}) with large dimensions, i.e., 
for large $K$, $L$, $n_t$, keeping $L/K$ fixed. 

\section{Detection Using BP on Markov Random Fields }
\label{sec3}
In this section, we present a detection algorithm based on message
passing on a MRF graphical model of the MIMO system model in 
(\ref{eqn1}) \cite{isi_icc2010}.

\subsection{Markov Random Fields}
\label{sec_3a}
\begin{figure}
\centering
\includegraphics[width=2.25in, height=1.30in]{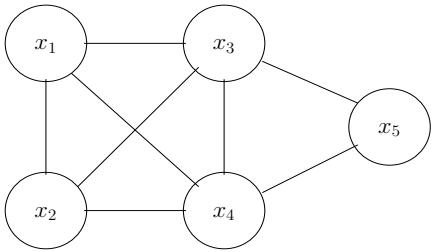}
\caption{An example of MRF.}
\label{fig2}
\end{figure}

An undirected graph is given by $G=(V,E)$, where $V$ is the set of nodes and
$E\subseteq \left \{(i,j): i,j\in V, i\neq j \right\}$ is the set
of undirected edges. An MRF is an undirected graph whose vertices are random
variables \cite{Pearl1988},\cite{frey}. The statistical dependency among
the variables are such that any variable is independent of all the other
variables, given its neighbors.
Usually, the variables in an MRF are constrained by a \textit{compatibility
function}, also known as a \textit{clique potential} in literature. A
\textit{clique} of an MRF is a fully connected sub-graph, i.e., it is a
subset $C \subseteq V$ such that $(i,j) \in E$ for all $i,j \in C$. A clique
is \textit{maximal} if it is not a strict subset of another clique. Therefore,
a maximal clique does not remain fully connected if any additional
vertex of the MRF is included in it. For example, in the MRF shown in Fig.
\ref{fig2}, $\left\{x_1,x_2,x_3,x_4\right\}$ and
$\left \{x_3,x_4,x_5\right\}$ are two maximal cliques.

Let there be $N_c$ maximal cliques in the MRF, and $\mathbf{x}_j$ be the
variables in maximal clique $j$. Let $\psi_j\left(\mathbf{x}_j\right)$ be
the clique potential of clique $j$. Then the joint distribution of the
variables is given by Hammersley-Clifford theorem \cite{griffeath}
\begin{eqnarray}
p\left(\mathbf{x}\right) & = & \frac{1}{Z} \prod_{j = 1}^{N_c} \psi_j\left(\mathbf{x}_j\right),
\end{eqnarray}
where $Z$ is a constant, also known as {\em partition function}, chosen to 
ensure the distribution is normalized.
In  Fig. \ref{fig2}, with two maximal cliques in the MRF, namely,
$\left\{x_1,x_2,x_3,x_4\right\}$ and $\left\{x_3,x_4,x_5\right\}$, the
joint probability distribution is given by
\begin{eqnarray}
p\left(\mathbf{x}\right) & = & \frac{1}{Z}\, \psi_1\left(x_1,x_2,x_3,x_4\right) \psi_2\left(x_3,x_4,x_5\right).
\end{eqnarray}

{\em Pairwise MRF:} An MRF is called a {\em pairwise} MRF if all the maximal 
cliques in the MRF are of size two. In this case, the clique potentials are 
all functions of two variables. The joint distribution in such a case takes 
the form \cite{merl}
\begin{eqnarray}
p\left(\mathbf{x}\right) & \propto & \Big(\prod_{(i,j)}\psi_{i,j}\left( x_i,x_j \right) \Big) \Big( \prod_{i}\phi_i\left(x_i\right) \Big),
\label{eqnz2}
\end{eqnarray}
where $\psi_{i,j}\left(x_i,x_j\right)$ is the clique potential between
nodes $x_i$ and $x_j$ denoting the statistical dependence between them,
and $\phi_{i}\left(x_i\right)$ is the self potential of node $x_i$.

\subsection{MRF of MIMO System} 
\begin{figure}
\centering
\includegraphics[width=2.25in, height=2.25in]{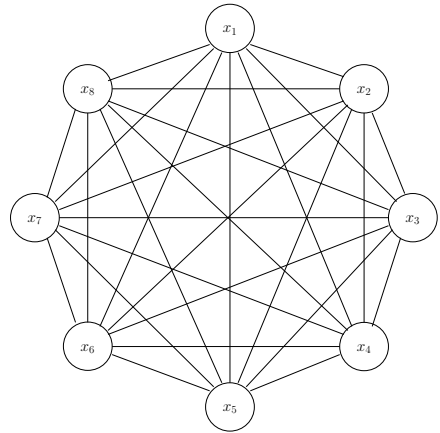}
\caption{Fully connected MRF of $8\times 8$ MIMO system.}
\label{fig2x}
\end{figure}

The MRF of a MIMO system is a fully connected graph. Figure \ref{fig2x} 
shows the MRF for a $8\times 8$ MIMO system. We get the MRF potentials 
for the MIMO system where the posterior probability function of the 
random vector ${\bf x}$, given ${\bf r}$ and ${\bf H}$, is of the 
form\footnote{In 
our detection problem, relative values of the distribution for various 
possibilities of $\mathbf{x}$ are adequate. So, we can omit the normalization 
constant $Z$, which is independent of ${\bf x}$, and replace the equality with 
proportionality in the distribution.}

\begin{eqnarray}
\hspace{-0mm}
p({\bf x}\mid {\bf r},{\bf H}) & \propto & \exp\Big(\frac{-1}{2\sigma^2}\|{\bf r}-{\bf H}{\bf x} \|^2 \Big) \exp\big(\ln p({\bf x})  \big) \nonumber \\
&=& \exp\Big(-\frac{1}{2\sigma^2}({\bf r}-{\bf H}{\bf x})^H ({\bf r}-{\bf H}{\bf x})\Big) \nonumber \\
& & \cdot \prod_{i} \exp\big(\ln p(x_i) \big) \nonumber \\
&\propto& \exp\Big(-\frac{1}{2\sigma^2}\big({\bf x}^H{\bf H}^H{\bf H}{\bf x}-2\Re\{{\bf x}^H{\bf H}^H{\bf r}\} \big) \Big) \nonumber \\
& & \cdot \prod_{i} \exp\big(\ln p(x_i) \big). 
\label{dist}
\end{eqnarray}
Now, defining
${\bf R} \Define \frac{1}{\sigma^2} {\bf H}^H{\bf H} \;$ and 
$\;{\bf z}\Define \frac{1}{\sigma^2} {\bf H}^H{\bf r}$,
we can write (\ref{dist}) as

{\small
\begin{eqnarray}
p({\bf x} \mid {\bf r},{\bf H}) & \propto & \exp\Big(-\sum_{i<j}\Re\{x_i^* R_{ij} x_j\}\Big) \nonumber \\
& & \cdot \exp \Big(\sum_{i}\Re\{x_i^*z_i\} \Big) \prod_{i} \exp\big(\ln p(x_i) \big) \nonumber \\
&\hspace{-4.6cm} = & \hspace{-2.6cm} \left( \prod_{i<j}\hspace{-0.5mm}\exp\hspace{-0.5mm}\Big(\hspace{-1mm}-\hspace{-1mm}x_i\Re\{R_{ij}\}x_j\Big) \hspace{-1mm} \right) \hspace{-1mm} \left( \prod_{i}\hspace{-0.5mm}\exp\hspace{-0.5mm}\Big(x_i\Re\{z_i\}+\ln p(x_i)\Big) \hspace{-1mm} \right)\hspace{-1mm},
\label{dist2}
\end{eqnarray}
}

\hspace{-5mm}
where $z_i$ and $R_{ij}$ are the elements of
${\bf z}$ and ${\bf R}$, respectively.
Comparing (\ref{dist2}) and (\ref{eqnz2}), we see that the MRF of the MIMO 
system has only pairwise interactions with the following potentials 
\begin{eqnarray}
\psi_{i,j}\left( x_i,x_j\right) & = & \exp\Big(\hspace{-1mm}-\hspace{-1mm}x_i\Re\{R_{ij}\}x_j\Big),  
\label{psix1} \\
\phi_i\left(x_i\right) & = & \exp\Big(x_i\Re\{z_i\}+\ln p(x_i)\Big). 
\label{phix1}
\end{eqnarray}

\subsection{Message Passing}
\label{sec_3c}
The values of $\psi$ and $\phi$ given by (\ref{psix1}) and (\ref{phix1})
define, respectively, the edge and self potentials of an undirected
graphical model to which message passing algorithms, such as belief
propagation (BP),
can be applied to compute the marginal probabilities of the variables.
BP attempts to estimate the marginal probabilities of all the variables
by way of passing messages between the local nodes.

A \textit{message} from node $j$ to node $i$ is denoted as
$m_{j,i}\left(x_i\right)$, and belief at node $i$ is denoted as
$\text{b}_i(x_i)$, $x_i \in \{\pm 1\}$. The $\text{b}_i(x_i)$ is
proportional to how likely $x_i$ was transmitted. On the other hand,
$m_{ji}(x_i)$ is proportional to how likely $x_j$ thinks
$x_i$ was transmitted. The belief at node $i$ is
\begin{eqnarray}
\text{b}_i\left(x_i\right) & \propto & \phi_i\left(x_i\right) \prod_{j \in \mathcal{N} \left(i\right)} m_{j,i}\left(x_i\right),
\label{eqnmrfbelief}
\end{eqnarray}
where $\mathcal{N}(i)$ denotes the neighboring nodes of node $i$,
and the messages are defined as \cite{merl}
\begin{eqnarray}
\hspace{-7mm}
m_{j,i}\left(x_i\right) & \hspace{-1mm} \propto & \hspace{-1mm} \sum_{x_j}\phi_j\left(x_j\right) \psi_{j,i}\left(x_j,x_i\right) \prod_{k\in \mathcal{N}\left(j\right) \setminus i} \hspace{-2mm} m_{k,j}\left(x_j\right).
\label{eqnmsgdefn}
\end{eqnarray}

Equation (\ref{eqnmsgdefn}) actually constitutes an iteration, as the
message is defined in terms of the other messages. So, BP
essentially involves computing the outgoing messages from
a node to each of its neighbors using the local joint compatibility
function and the incoming messages and transmitting them. The algorithm
terminates after a fixed number of iterations.

\subsection{Improvement through Damping}
\label{sec_3d}
In systems characterized by fully/highly connected graphical models,
BP based algorithms may fail to converge, and if they do converge, the
estimated marginals may be far from exact \cite{mooij3},\cite{mooij2}.
It may be expected that BP might perform poorly in MIMO graphs due
to the high density of connections. However, several methods are known
in the literature, including {\em double loop methods} 
\cite{Heskes},\cite{yuille} and {\em damping} \cite{damp},\cite{loopybp6}
which can be applied to improve things if BP does not converge (or
converges too slowly). In this paper, we consider damping methods.

In \cite{damp}, Pretti proposed a modified version of BP with
over-relaxed BP dynamics. At each step of the algorithm, the evaluation
of messages is taken to be a weighted average between the old estimate
and the new estimate. The weighted average could either be applied to the
messages (resulting in {\em message damped BP}) or to the estimate of
the probability distribution/beliefs of the variables
({\em probability/belief damped BP}), or to both messages and beliefs
({\em hybrid damped BP}). It is shown, in \cite{damp}, that the
probability damped BP can be derived as a limit case in which the
double-loop algorithm becomes a single-loop one.

{\em Message Damped BP:}
Denoting ${\widetilde{m}}_{i,j}^{(t)}(x_j)$ as the updated message in
iteration $t$ obtained by message passing, the new message from node
$i$ to node $j$ in iteration $t$, denoted by $m_{i,j}^{(t)}(x_j)$, is
computed as a convex combination of the old message and the updated
message as
\begin{eqnarray}
\hspace{-7.5mm}
{\widetilde{m}}_{i,j}^{(t)}(x_j) & \hspace{-2mm} \propto & \hspace{-2mm}
 \sum_{x_i}\phi_i\left(x_i\right) \psi_{i,j}\left(x_i,x_j\right) \hspace{-1mm} \prod_{k\in \mathcal{N}\left(i\right) \setminus j}\hspace{-3mm} m_{k,i}^{(t-1)}\left(x_i\right),
\label{eqn_m1}
\end{eqnarray}
\begin{eqnarray}
\hspace{-4mm}
m_{i,j}^{(t)}(x_j) & \hspace{-1mm} = & \hspace{-1mm} \alpha_m\, m_{i,j}^{(t-1)}(x_j) + (1-\alpha_m)\,{\widetilde{m}}_{i,j}^{(t)}(x_j),
\label{eqn_m}
\end{eqnarray}
where $\alpha_m \in [0,1)$ is referred as the {\em message damping
factor}.

{\em Belief Damping:}
Instead of damping the messages in each iteration, the beliefs of
the variables can be computed in each iteration as a weighted
average, as

\begin{eqnarray}
\widetilde{\text{b}}_i^{(t)}(x_i) & \propto & \phi_i(x_i) \prod_{j \in \mathcal{N}(i)} m_{j,i}^{(t)}(x_i),
\label{eqn_b1}
\end{eqnarray}
\begin{eqnarray}
\text{b}_i^{(t)}(x_i) & = & \alpha_b\, \text{b}_i^{(t-1)}(x_i) + (1-\alpha_b)\, \widetilde{\text{b}}_i^{(t)}(x_i),
\label{eqn_b}
\end{eqnarray}
where $\alpha_b \in [0,1)$ is referred to as the {\em belief damping factor}.

{\em Hybrid Damping:} As a more general damping strategy, we can update
both the messages as well as the beliefs according to (\ref{eqn_m}) and
(\ref{eqn_b}), respectively, in each iteration. Different combinations
of $(\alpha_m,\alpha_b)$ values
specializes to different strategies; for e.g.,
{\small $(\alpha_m = \alpha_b = 0)$} corresponds to Undamped BP,
{\small $(\alpha_m \neq 0, \alpha_b = 0)$} corresponds to Message damped BP,
{\small $(\alpha_m = 0, \alpha_b \neq 0)$} corresponds to Belief damped BP,
and {\small $(\alpha_m \neq 0, \alpha_b \neq 0)$} corresponds to Hybrid
damped BP. 

The proposed BP algorithm employing damping is listed in Table \ref{table2}. 

\subsection{Computation Complexity}
\label{sec_3e}
The per-symbol complexity of calculating messages and beliefs in a
single BP iteration is $O(K^2n_t^2)$ and $O(Kn_t)$, respectively.
Likewise, the per-symbol complexity of computing $\phi$ and $\psi$
is $O(1)$ and $O(Kn_t)$, respectively. The computation of ${\bf z}$
can be carried out with $O(Kn_r)$ per-symbol complexity. The computation 
of ${\bf R}$ involves computation of {\small ${\bf H}^H{\bf H}$}, which 
involves three operations: $i)$ computation of ${\mathbf G}$, $ii)$
calculation of {\small ${\mathbf G}^H{\mathbf G}$}, and $iii)$ 
multiplication of ${\mathbf F}^H$ and ${\mathbf F}$ with 
${\mathbf G}^H{\mathbf G}$. The computation $i)$ involves $K$-point 
FFT of matrices $H_l$, $l=0,\cdots,L-1$, each $H_l$ of dimension 
$n_r \times n_t$. The complexity associated with this operation is 
$O(n_tn_rK\log_2K)$. The total number of symbols transmitted is $Kn_t$. 
So, the per-symbol complexity is $O(n_r\log_2K)$. The computation $ii)$ 
involves the calculation of ${{\mathbf G}_i}^H{\mathbf G}_i$ for 
$i=0,\cdots,K-1$. The computation of each ${{\mathbf G}_i}^H{\mathbf G}_i$ 
has complexity $O(n_t^3)$. Due to block-diagonal structure of $\mathbf G$, 
$K$ such computations can be done in $O(Kn_t^3)$ complexity, leading to
a per-symbol complexity of $O(n_t^2)$. Likewise, due to the block-symmetric 
structure of ${\mathbf F}$, the per-symbol complexity corresponding to 
computation $iii)$ is $O(Kn_t^2)$. Since the number of BP iterations is 
much less than $Kn_t$, the overall per-symbol complexity is of the 
proposed MRF based BP detection algorithm is given by $O(K^2n_t^2)$, 
which scales well for large $Kn_t$.

\begin{table}
\begin{center}
\begin{tabular}{|l|}
\hline \\
\hspace{3mm}
{\em Initialization} \\

1.  $m_{i,j}^{(0)}(x_j) = \text{b}_i^{(0)}(x_i) = 0.5$, \\

    $\,\,$ $p(x_i=1)=p(x_i=-1)=0.5$, $\,$ $\forall i,j=1,\cdots,Kn_t$ \\

2.  ${\widetilde{m}}_{i,j}^{(0)}(x_j) = \widetilde{\text{b}}_i^{(0)}(x_i)=0.5$, $\,$ $\forall i,j=1,\cdots,Kn_t$ \\

3.  ${\bf z} = \frac{1}{\sigma^2}\,{\bf H}^{H}{\bf r}$; \,\,\,
    $\mathbf{R} = \frac{1}{\sigma^2}\,{\bf H}^{H}{\bf H}$  \\

4. $\,\,$ for $i$ = 1 to $Kn_t$  \\

5.  $\,\,\,\,\,\, \phi_i(x_i) = \exp\big( x_i\Re\left\{z_i\right\} + \ln(p(x_i)) \big)$ \\

6. $\,\,$ end for \\

7. $\,$ for $i$ = 1 to $Kn_t$ \\

8. $\,\,\,\,\,$for $j$ = 1 to $Kn_t$, $\,\,\,j \neq i$ \\

9. $\,\,\,\,\,\,\,\, \psi_{i,j}\left( x_i,x_j\right) = \exp\big(- x_i\Re\left\{R_{i,j}\right\}x_j \big)$ \\

10. $\,\,\,$end for \\

11. end for \\

\hspace{3mm}{\em Iterative Update of Messages and Beliefs} \\

12. for $t$ = 1 to $num\_iter$ \\

\hspace{5mm}{\em Damped Message Calculation} \\

13. $\,\,\,$for $i$ = 1 to $Kn_t$ \\

14. $\,\,\,\,\,\,$for $j$ = 1 to $Kn_t, \,\,\, j \neq i$ \\

15. $\,\,\,\,\,\,\,\,\, {\widetilde{m}}_{i,j}^{(t)}(x_j) \propto  \sum_{x_i}\phi_i(x_i) \psi_{i,j}(x_i,x_j)$ \\

    \hspace{4cm}$\cdot \prod_{k\in \mathcal{N}(i) \setminus j} m_{k,i}^{(t-1)}(x_i)$ \\

16. $\,\,\,\,\,\,\,\,\, m_{i,j}^{(t)}(x_j) = \alpha_m\, m_{i,j}^{(t-1)}(x_j) + (1-\alpha_m)\,{\widetilde{m}}_{i,j}^{(t)}(x_j)$ \\

17. $\,\,\,\,\,\,$end for \\

18. $\,\,\,$end for \\

\hspace{5mm}{\em Damped Belief Calculation}\\

19. $\,\,\,$for $i$ = 1 to $Kn_t$ \\

20. $\,\,\,\,\,\,$ $\widetilde{\text{b}}_i^{(t)}(x_i) \propto \phi_i(x_i) \prod_{j \in \mathcal{N}(i)} m_{j,i}^{(t)}(x_i)$ \\
21. $\,\,\,\,\,\,$ $\text{b}_i^{(t)}(x_i) \propto \alpha_b\, \text{b}_i^{(t-1)}(x_i) + (1-\alpha_b)\,\widetilde{\text{b}}_i^{(t)}(x_i)$ \\

22. $\,\,\,\,$end for \\

23. end for; \hspace{5mm} End of for loop starting at line 12 \\

24. ${\widehat x}_i={\mbox{arg max}\atop{{x_i\in \{\pm 1\}}}} \,\, \text{b}_i^{(num\_iter)}\left(x_i\right), \,\,\, \forall \, i=1,\cdots,Kn_t$\\

25. Terminate \\
\hline
\end{tabular}
\caption{Proposed MRF Based BP Detector/Equalizer Algorithm.}
\label{table2}
\end{center}
\vspace{-8mm}
\end{table}

\subsection{Simulation Results }
\label{sec_3f}
In this section, we present the simulated BER performance of the
proposed MRF BP detection algorithm. 

\begin{figure}
\hspace{-2mm}
\includegraphics[width=3.75in, height=2.90in]{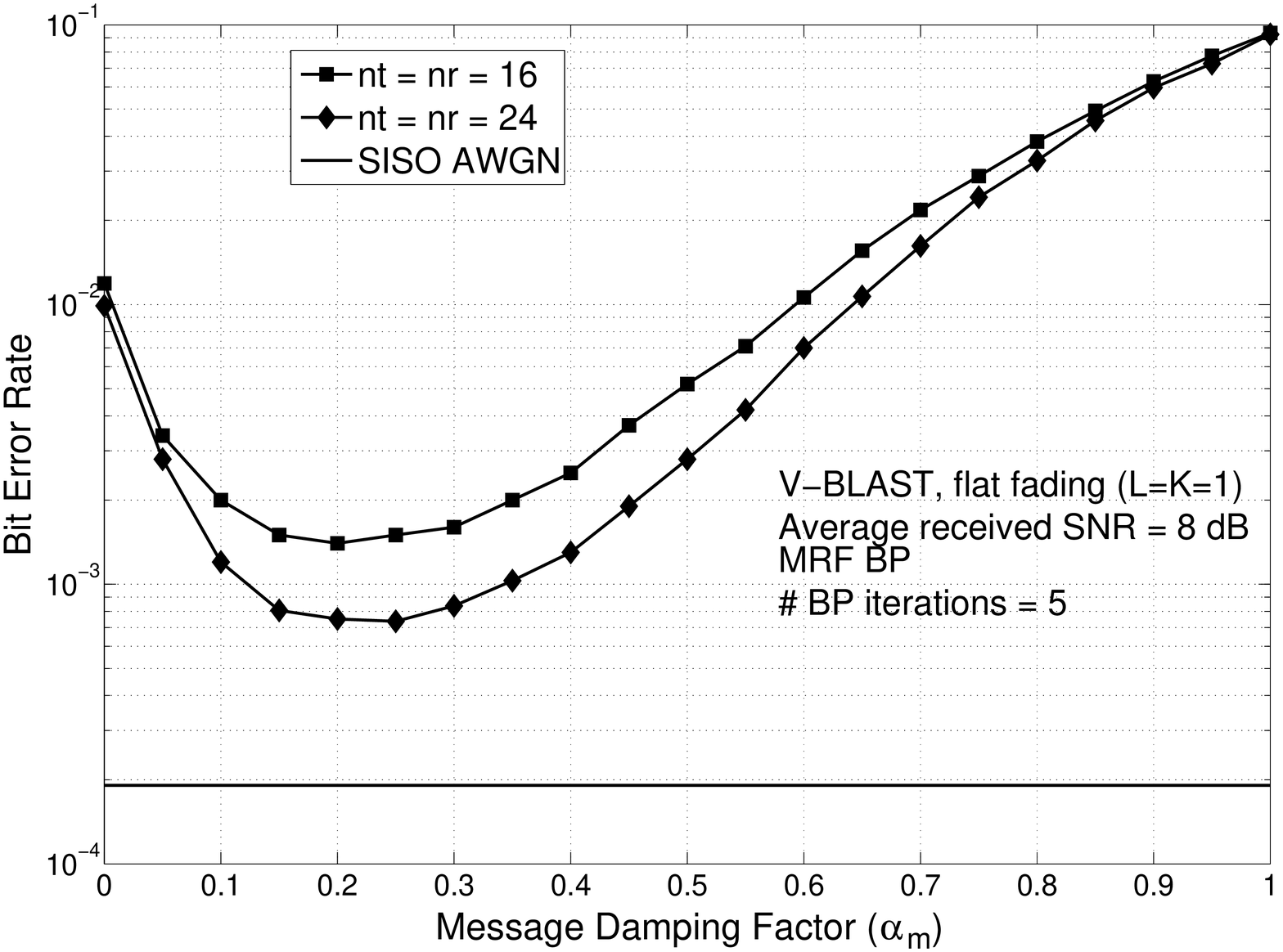}
\vspace{-6mm}
\caption{BER performance of the MRF BP algorithm as a 
function of message damping factor, $\alpha_m$, in V-BLAST MIMO with 
$n_t=n_r=16,24$ on flat fading ($L=K=1$) at 8 dB SNR. \# BP iterations=5.}
\label{fig4}
\end{figure}

\begin{figure}
\hspace{-2mm}
\includegraphics[width=3.75in, height=2.90in]{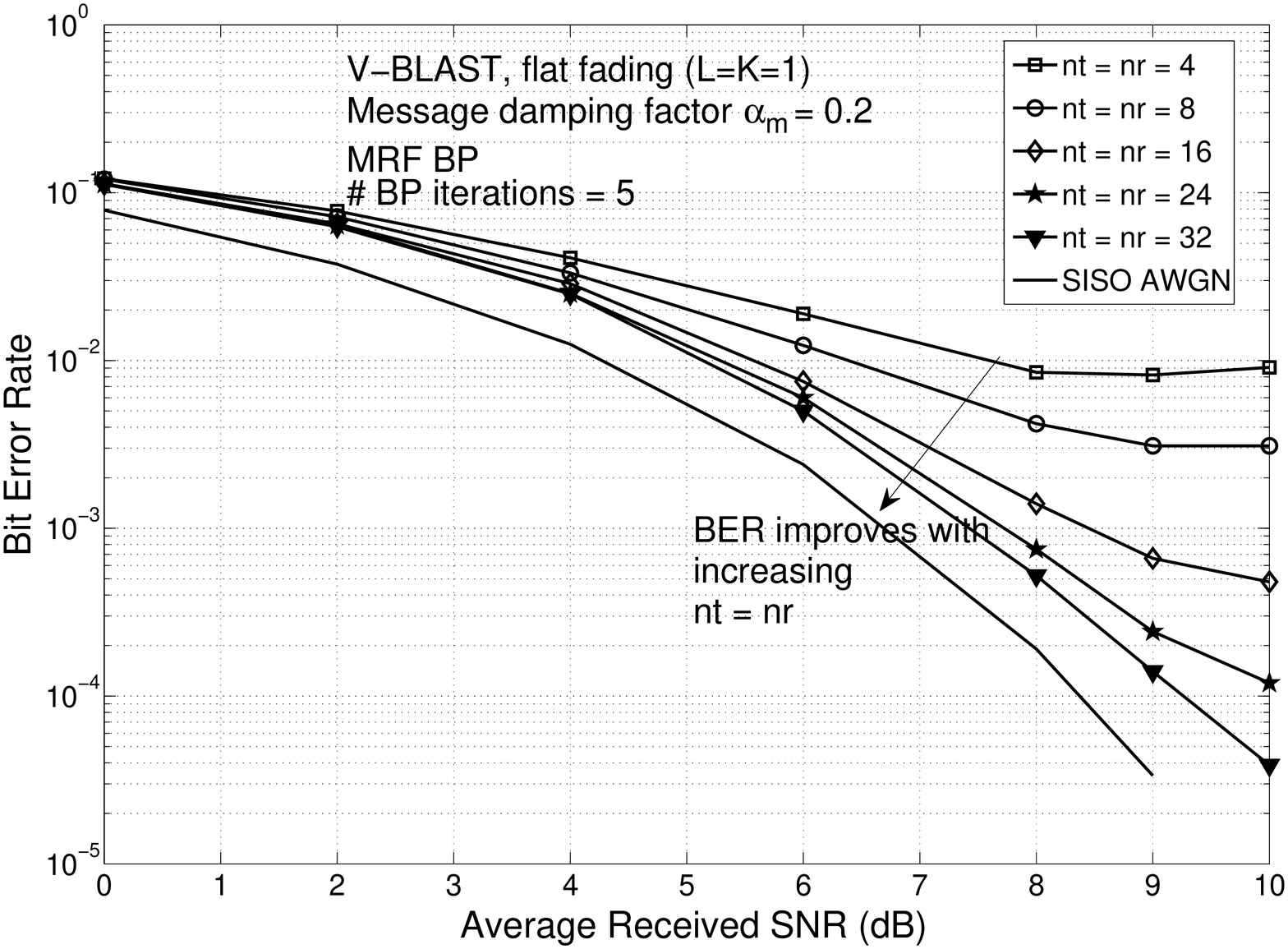}
\vspace{-6mm}
\caption{BER performance of the MRF BP algorithm as a 
function of SNR in V-BLAST MIMO for different $n_t=n_r$ on flat fading 
($L=K=1$) with message damping, $\alpha_m=0.2$, and \# BP iterations = 5.}
\label{fig5}
\end{figure}

\begin{figure}
\hspace{-2mm}
\includegraphics[width=3.75in, height=2.90in]{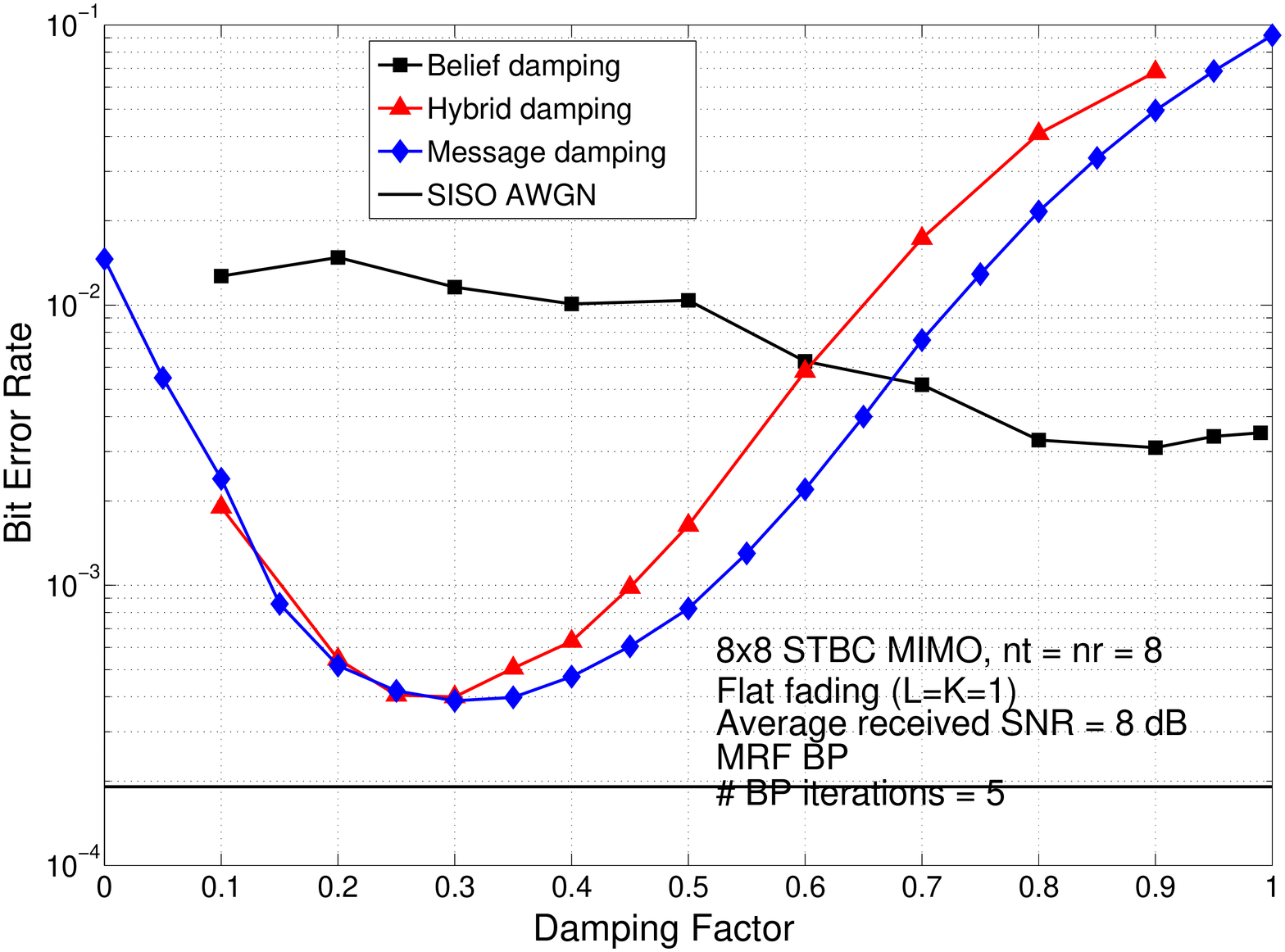}
\vspace{-6mm}
\caption{Effect of message, belief, and hybrid damping on the BER
performance of $8\times 8$ STBC from CDA with $t=e^{{\bf j}}$,
$\delta=e^{\sqrt{5}\,{\bf j}}$, $n_t=n_r=8$ on flat fading
($L=K=1$) at 8 dB SNR. MRF BP,
\# BP iterations = 5, $\alpha_m=\alpha_b$ for hybrid damping.}
\label{fig6}
\end{figure}

{\em Performance in Flat-Fading with Large $n_t$:}
In Figs. \ref{fig4} to \ref{fig6}, we illustrate the `large-dimension 
behavior' of the algorithm and the effect of damping for large number 
(tens) of transmit and receive antennas with BPSK modulation on flat 
fading channels (i.e., $L=K=1$). The number of BP iterations is 5. 
Figure \ref{fig4} shows the variation of the achieved BER as a function of 
the message damping factor, $\alpha_m$, in $16\times 16$ and $24\times 24$ 
V-BLAST MIMO systems at an average received SNR per receive antenna, $\gamma$, 
of 8 dB. Note that $\alpha_m=0$ corresponds to the case of undamped BP. It 
can be observed from Fig. \ref{fig4} that, depending on the choice of the 
value of $\alpha_m$, message damping can significantly improve the BER 
performance of the BP algorithm. There is an optimum value of $\alpha_m$ at 
which the BER improvement over no damping case is maximum. For the chosen 
set of system parameters in Fig. \ref{fig4}, the optimum value of $\alpha_m$ 
is observed to be about 0.2. For this optimum value of $\alpha_m=0.2$, it is 
observed that about an order of BER improvement is achieved with message 
damping compared to that without damping. From Fig. \ref{fig4}, it 
can further be seen that the performance improves for increasing $n_t=n_r$ 
(i.e., performance of the $n_t=n_r=24$ system is better that of the 
$n_t=n_r=16$ system). This shows that the algorithm exhibits `large-dimension 
behavior,' where the BER performance moves closer towards unfaded SISO AWGN 
performance when $n_t=n_r$ is increased from 16 to 24. This large-dimension 
behavior is illustrated even more clearly in Fig. \ref{fig5}, where we plot 
the BER performance of V-BLAST MIMO as a function of SNR for different 
$n_t=n_r=4,8,16,24$ and $32$ for $\alpha_m=0.2$.

In Fig. \ref{fig6}, we present a comparison of the BER performance
achieved using message damping, belief damping and hybrid damping based
BP detection of $8\times 8$ non-orthogonal space-time block code (STBC) 
from cyclic division algebra (CDA) with
$t=e^{{\bf j}}$, $\delta=e^{\sqrt{5}\,{\bf j}}$ \cite{bsr} at 8 dB SNR.
In this type of STBC, each STBC is a $n_t\times p$ square matrix with 
$n_t$ transmit antennas and $p=n_t$ time slots constructed using $n_t^2$ 
symbols, which results in $n_t^2$ dimensions and $n_t$ symbols per channel 
use. For message damping and belief damping, $\alpha_m$ and $\alpha_b$ are
varied in the range 0 to 1. For hybrid damping, we set $\alpha_m=\alpha_b$
and varied it in the range 0 to 1. From Fig. \ref{fig6}, it can be seen that 
$i)$ with damping, there is an optimum value of the damping factor at which 
the BER performance is the best (e.g., for message damping, the optimum 
damping factor is about 0.3 in Fig. \ref{fig6}), $ii)$ message damping 
performs better than belief damping for small values of the damping factor, 
whereas belief damping performs better at high values of the damping factor; 
however, over the entire range of the damping factor, the best performance 
of message damping is significantly better than the best performance of 
belief damping, and $iii)$ for the chosen condition of $\alpha_m=\alpha_b$, 
hybrid damping performance is similar to that of message damping; however, 
$\alpha_m$ and $\alpha_b$ in hybrid damping can be jointly optimized to
further improve the performance.

\begin{figure}
\hspace{-3mm}
\includegraphics[width=3.75in, height=2.90in]{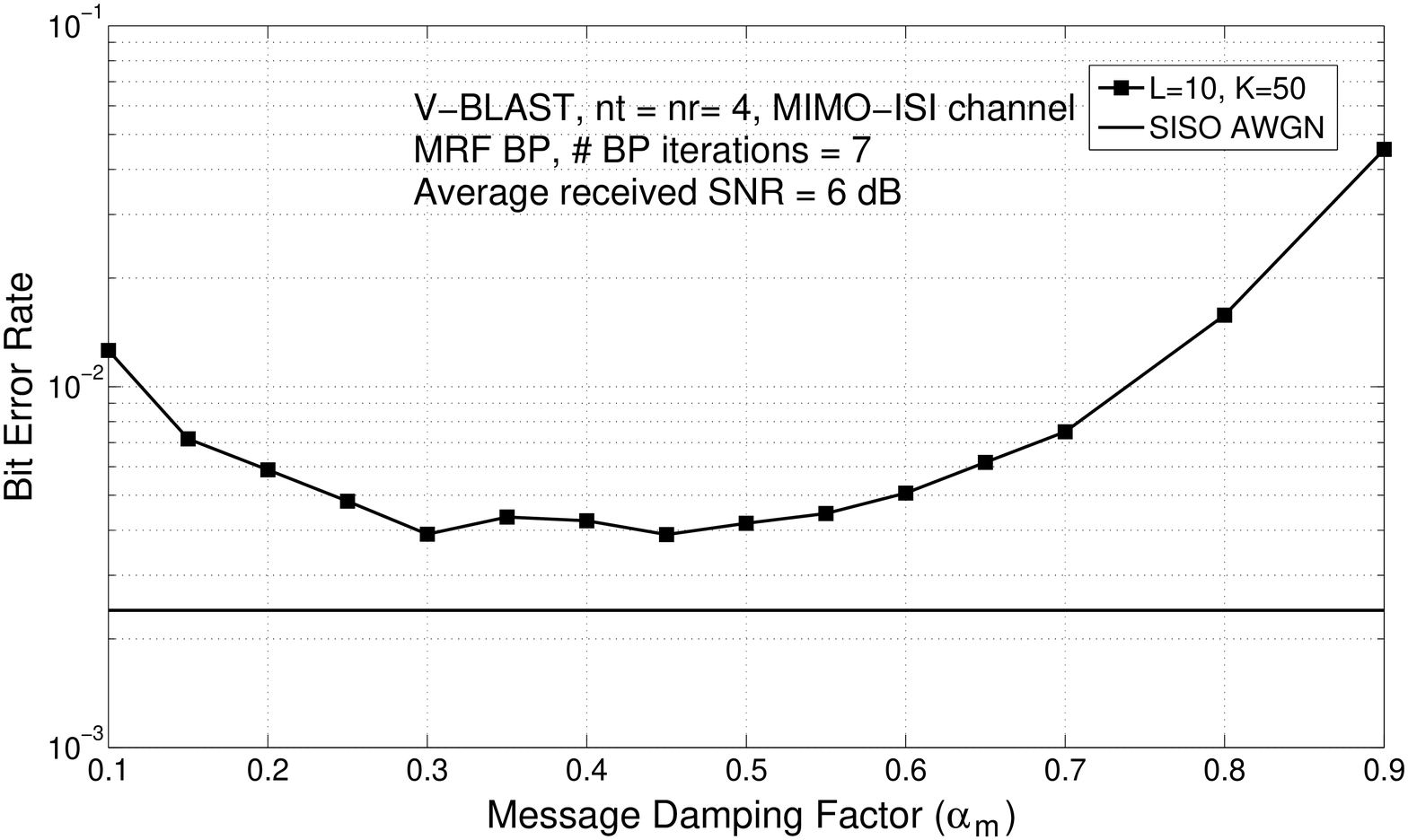}
\vspace{-5mm}
\caption{BER performance of the MRF BP algorithm as
a function of the message damping factor, $\alpha_m$, in MIMO-ISI channels.
$n_t=n_r=4$, $[L=10, K=50]$, uniform power delay profile, average received 
SNR = 6 dB, \# BP iterations = 7.}
\label{fig7}
\end{figure}

\begin{figure}
\hspace{-3mm}
\includegraphics[width=3.75in, height=2.90in]{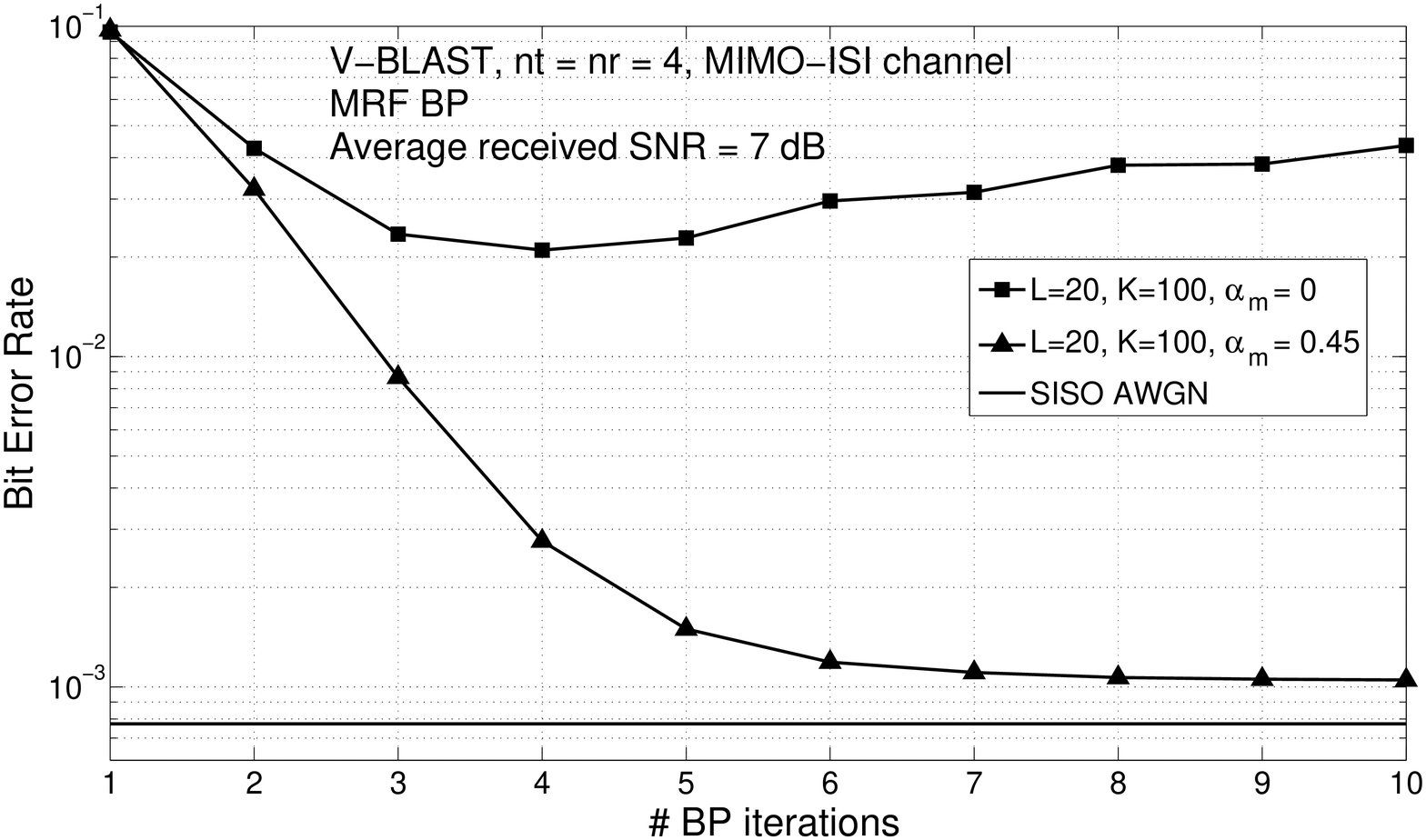}
\vspace{-5mm}
\caption{Comparison of the BER performance of message damped and undamped 
MRF BP detector/equalizer as a function of number of BP iterations in 
MIMO-ISI channels. $n_t=n_r=4$, $[L=20, K=100]$, uniform power delay profile,
average received SNR = 7 dB, $\alpha_m=0$ (undamped), $\alpha_m= 0.45$ 
(damped). }
\label{fig8}
\vspace{-3mm}
\end{figure}

{\em Performance in MIMO-ISI Channels with Large $Kn_t$:}
In Fig. \ref{fig7}, we explore the effect of message damping on the
BER performance of the MRF based BP detector/equalizer in MIMO-ISI 
channels. In all the simulations of MIMO-ISI channels, 
we have taken uniform power delay profile (i.e., all the $L$ paths are
assumed to have equal energy).
Figure \ref{fig7} shows the variation of the achieved BER as a function 
of the message damping factor, $\alpha_m$, 
for $n_t=n_r=4$, BPSK, $[L=10, K=50]$, at an average received SNR of 6 dB. 
The total number of dimensions, $Kn_t=200$. The number of BP iterations used 
is 7. From Fig. \ref{fig7}, it is can be seen that
damping can significantly improve the BER performance of the BP algorithm. 
For the chosen set of system parameters in Fig. \ref{fig7}, the optimum 
value of $\alpha_m$ is observed to be about 0.45, which gives about an 
order of BER improvement. This point of the benefit of damping in terms of 
BER performance (and also in terms of convergence) is even more clearly 
brought out in Fig. \ref{fig8}, where we have compared the BER performance 
without damping ($\alpha_m=0$) and with damping ($\alpha_m=0.45$) for 
$[L=20, K=100]$ at an SNR of 7 dB as a function of the number of BP 
iterations. It is interesting to see that without damping (i.e., with 
$\alpha_m=0$), the algorithm indeed shows `divergence' behavior, i.e., 
BER increases as number of iterations is increased beyond 4. Such divergence 
behavior is effectively removed by damping, as can be seen from the BER 
performance achieved with $\alpha_m=0.45$. Indeed the algorithm with 
damping ($\alpha_m=0.45$) is seen to be converge smoothly. It is also 
interesting to note that the algorithm converges to a BER which is quite 
close to the unfaded SISO AWGN BER (BER on SISO AWGN at 7 dB SNR is about 
$7.8\times 10^{-4}$ and the converged BER using damped BP is about 
$1\times 10^{-3}$). This illustrates the potential of damping in improving 
BER performance and convergence of the algorithm when employed for 
detection/equalization in the considered MIMO system on severely delay spread 
frequency-selective channels (e.g., $L=20$). It is also noted that damping 
(as per Eqn. (\ref{eqn_m})) does not increase the order of complexity of the 
algorithm without damping; the order of complexity without and with damping 
remains the same.

\begin{figure}
\hspace{-3mm}
\includegraphics[width=3.75in, height=2.90in]{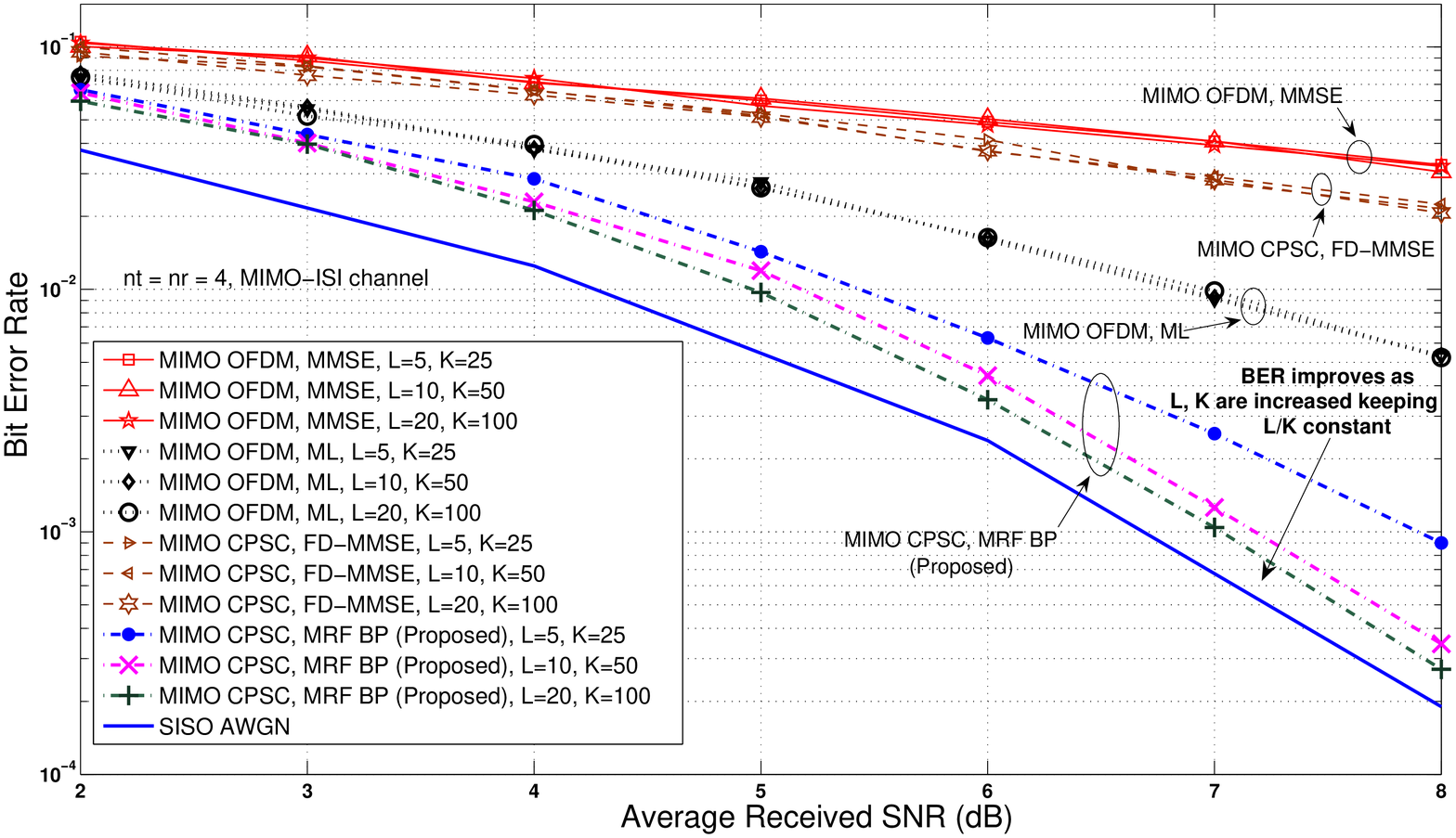}
\vspace{-7mm}
\caption{BER performance of message damped MRF BP detector/equalizer 
as a function of average received SNR in MIMO-ISI channels with $n_t=n_r=4$ 
for different values of $L$ and $K$ keeping $L/K$ constant: $[L=5, K=25]$, 
$[L=10,K=50]$, and $[L=20, K=100]$. Uniform power delay profile. \# BP 
iterations = 10, $\alpha_m=0.45$.}
\label{fig9}
\vspace{-6mm}
\end{figure}

{\em Comparison with MIMO-OFDM Performance:}
In Fig. \ref{fig9}, we present a performance comparison between the 
considered MIMO-CPSC scheme and a MIMO-OFDM scheme for the same system/channel
parameters in both cases; for $n_t=n_r=4$ and following combinations of 
$L$ and $K$: $[L=5, K=25]$, $[L=10,K=50]$, $[L=20, K=100]$. For MIMO-CPSC, two 
detection schemes are considered: FD-MMSE and proposed MRF BP. For the 
MRF BP, number of BP iterations used is 10 and the value of $\alpha_m$ 
used is 0.45. For MIMO-OFDM, two detection schemes, namely, MMSE and ML 
detection on each subcarrier are considered.
We have also plotted the unfaded SISO AWGN performance that serves as a 
lower bound on the optimum detection performance. The following observations
can be made from Fig. \ref{fig9}: $i)$ MIMO-OFDM with MMSE detection performs 
the worst among all the considered system/detection configurations, $ii)$ 
MIMO-CPSC with FD-MMSE performs better than MIMO-OFDM with MMSE (this better 
performance in CPSC is in line with other reported comparisons between 
OFDM and CPSC, e.g., \cite{cpsc1},\cite{cpsc2},\cite{cpsc3}), $iii)$ at the 
expense of increased detection complexity, MIMO-OFDM with ML detection 
performs better than both MIMO-OFDM with MMSE and MIMO-CPSC with FD-MMSE, 
and $iv)$ more interestingly, MIMO-CPSC with the low-complexity MRF BP 
detection significantly outperforms MIMO-OFDM even with ML detection. 
Indeed, the performance of the MIMO-CPSC with MRF BP detection gets 
increasingly closer to the SISO AGWN performance for increasing $L$, 
$K$, keeping $L/K$ constant. For example, the gap between the MRF BP 
performance and the SISO AWGN performance is only about 0.25 dB for $L=20$ 
at a BER of $10^{-3}$. This illustrates the ability of the MRF BP algorithm 
to achieve near-optimal performance for severely delay spread MIMO-ISI 
channels (i.e., large $L$) as witnessed in UWB systems.

\section{Detection using BP on Factor Graphs with Gaussian Approximation 
of Interference }
\label{sec4}
In this section, we present another low-complexity algorithm based on BP 
for detection in large-dimension MIMO-ISI channels. The graphical model 
employed here is factor graphs. A key idea in the proposed factor graph 
approach which enables to achieve low-complexity is the Gaussian 
approximation of interference (GAI) in the system.

\begin{figure*}
\begin{center}
\subfigure[]{
\includegraphics[width=2.5in,height=2.00in]{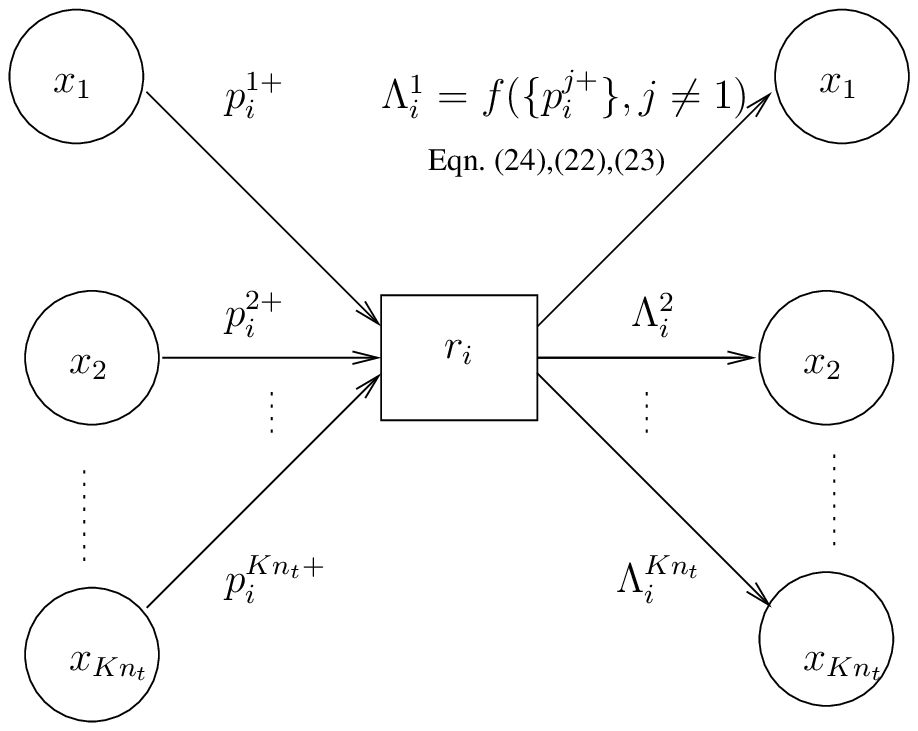}}
\hspace{10mm}
\subfigure[]{
\includegraphics[width=2.5in,height=2.00in]{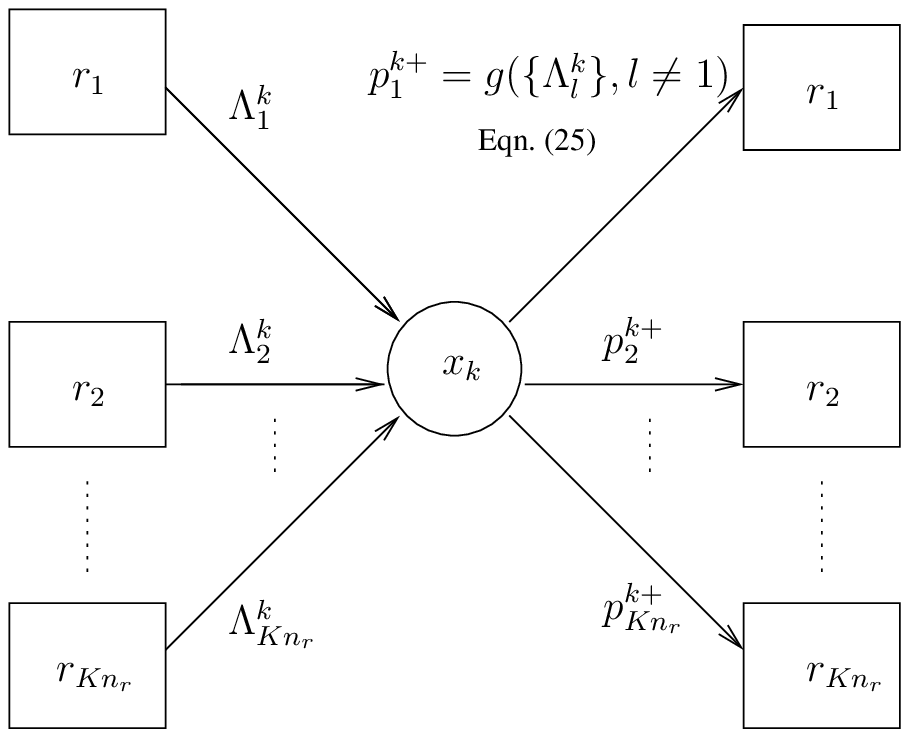}}
\caption{\label{fig10}
Message passing between variable nodes and observation nodes.}
\end{center}
\end{figure*}

Consider the MIMO system model in (\ref{eqn1}). We will treat each entry 
of the observation vector ${\bf r}$ as a function node (observation node)
in a factor graph, and each transmitted symbol as a variable node. 
The received signal $r_i$ can be written as
\begin{eqnarray}
r_i & = & \sum_{j=1}^{Kn_t}h_{ij} x_j + v_i \nonumber \\
    & = & h_{ik}x_k + \underbrace{\sum_{j=1,j\ne k}^{Kn_t}h_{ij} x_j}_{Interference} + \,\, v_i.
\label{model1}
\end{eqnarray}
When computing the message from the $i$th observation node to
the $k$th variable node, we make the following Gaussian approximation
of the interference:
\begin{eqnarray}
r_i & = & h_{ik}x_k + \underbrace{\sum_{j=1,j\ne k}^{Kn_t}h_{ij} x_j + v_i}_{\Define \,\, z_{ik} \, 
},
\label{model2}
\end{eqnarray}
where the interference plus noise term, $z_{ik}$, is modeled as
$\mathbb{C}{\cal N}(\mu_{z_{ik}},\sigma^2_{z_{ik}})$ with
\begin{eqnarray}
\mu_{z_{ik}} & = & 
\sum_{j=1,j\ne k}^{Kn_t}h_{ij}\mathbb{E}(x_j), 
\label{mu_z}
\end{eqnarray}
\begin{eqnarray}
\sigma^2_{z_{ik}} & = & 
\sum_{j=1,j\ne k}^{Kn_t}|h_{ij}|^2 \, \mbox{Var}(x_j) + \sigma^2. 
\label{sigma2_z}
\end{eqnarray}
For BPSK signaling, the log-likelihood ratio (LLR) of the symbol
$x_k \in \{+1,-1\}$ at observation node $i$, denoted by
$\Lambda_{i}^{k}$, can be written as
\begin{eqnarray}
\Lambda_{i}^{k} & = & \log\frac{p(r_i|{\bf H},x_k=1)}{p(r_i|{\bf H},x_k=-1)} \nonumber \\
& = & \frac{4}{\sigma_{z_{ik}}^2} \Re\left(h_{ik}^*(r_i-\mu_{z_{ik}})\right).
\label{LLR}
\end{eqnarray}

The LLR values computed at the observation nodes are passed to the variable 
nodes (Fig. \ref{fig10}a). Using these LLRs, the variable nodes compute 
the probabilities
\begin{eqnarray}
\hspace{-6mm}
p_i^{k+} & \Define & p_i(x_k=+1|{\bf r}) \nonumber \\
& = & \frac{\mbox{exp}(\sum_{l=1, l\neq i}^{Kn_r}\Lambda_l^k)}{1 + \mbox{exp}(\sum_{l=1, l\neq i}^{Kn_r}\Lambda_l^k)}, 
\label{prob}
\end{eqnarray}
and pass them back to the observation nodes (Fig. \ref{fig10}b). This message
passing is carried out for a certain number of iterations. Messages can be
damped as described in Section \ref{sec_3d} and then passed. Finally, $x_k$
is detected as
\begin{eqnarray}
\widehat{x}_k & = & \mbox{sgn}\Big(\sum_{i=1}^{Kn_r}\Lambda_i^k \Big).
\end{eqnarray}
Note that approximating the interference as Gaussian greatly simplifies
the computation of messages (as can be seen from the complexity discussion 
in the following subsection.) 

\subsection{Computation Complexity}
\label{sec_4a}
The computation complexity of the FG-GAI BP algorithm in the above 
involves $i)$ LLR calculations
at the observation nodes as per (\ref{LLR}), which has $O(K^2n_tn_r)$
complexity, and $ii)$ calculation of probabilities at variable nodes as
per (\ref{prob}), which also requires $O(K^2n_tn_r)$ complexity\footnote{A
naive implementation of (\ref{LLR}) would require a summation over 
$Kn_t-1$ variable nodes for each message, amounting to a complexity of
order $O(K^3 n_t^2 n_r)$. However, the summation over $Kn_t-1$ variables
in (\ref{mu_z}) can be written in the form
$\sum_{j=1}^{Kn_t}h_{ij}\mathbb{E}(x_j) - h_{ik}\mathbb{E}(x_k)$, 
where the computation of the full summation from $j=1$ to $Kn_t$ 
(which is independent of the variable index $k$) requires $Kn_t-1$ 
additions. In addition, one subtraction operation for each $k$ is
required. The makes the complexity order for computing (\ref{mu_z})
to be only $O(K^2n_tn_r)$. A similar argument holds for computation of
the variance in (\ref{sigma2_z}), and hence the complexity of computing
the LLR in (\ref{LLR}) becomes $O(K^2n_tn_r)$. Likewise, a similar
rewriting of the summation in (\ref{prob}) leads to a complexity of
$O(K^2n_tn_r)$.}. Hence, the overall complexity of the algorithm is 
$O(K^2n_tn_r)$ for detecting $Kn_t$ transmitted symbols. So the per-symbol 
complexity is just $O(Kn_t)$ for $n_t=n_r$. Note that this complexity is 
one order less than that of the MRF based approach in the previous section. 
Because of its linear complexity in $K$ and $n_t$, the proposed FG approach 
with GAI is quite attractive for detection in large-dimension MIMO-ISI 
channels. In addition, the BER performance achieved by the algorithm in 
large dimensions is very good (as shown in the BER performance results in 
the following subsection).  

\subsection{Simulation Results}
\label{sec_4b}
\begin{figure}
\hspace{-3mm}
\includegraphics[width=3.75in, height=2.90in]{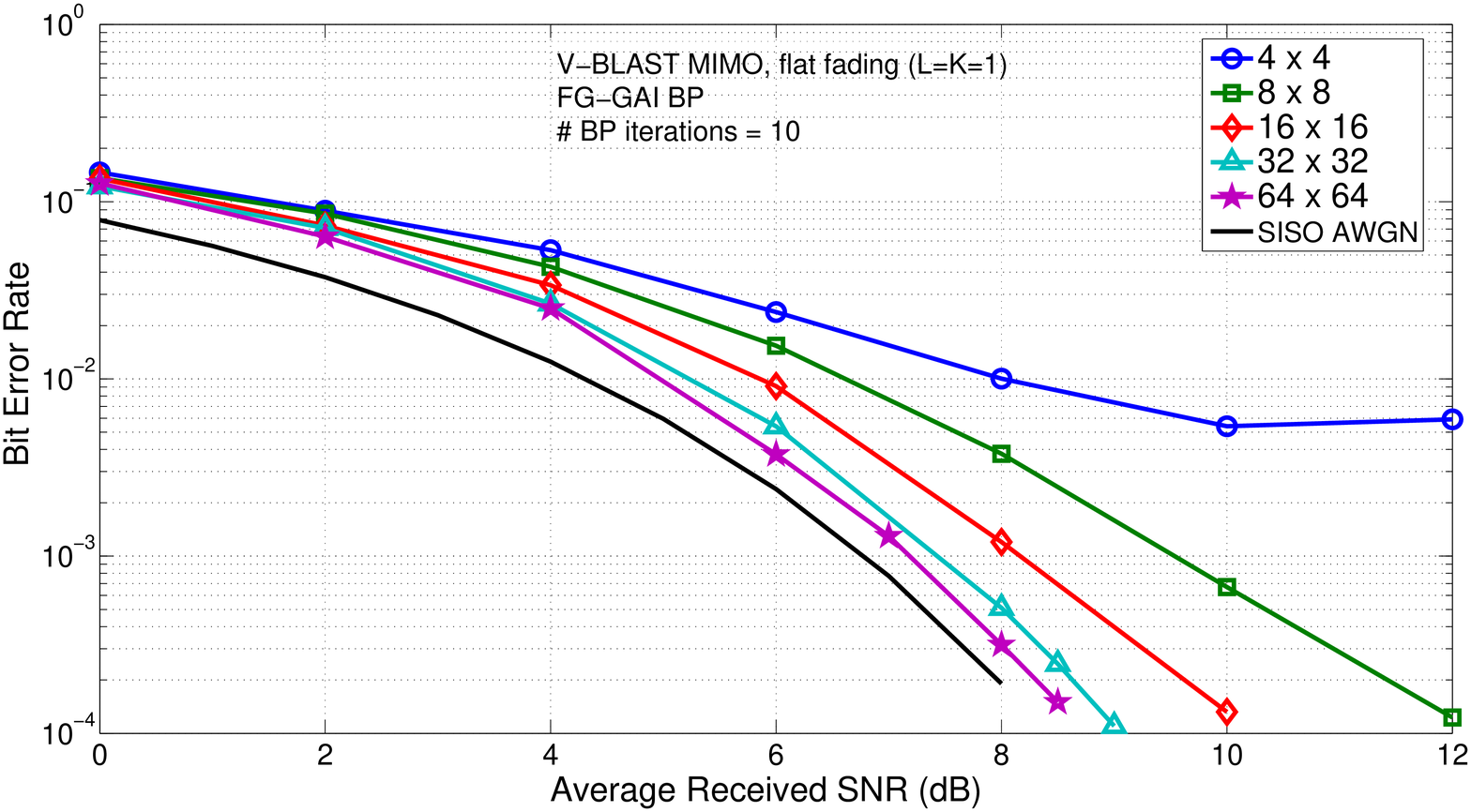}
\vspace{-6mm}
\caption{BER performance of the FG-GAI BP algorithm in V-BLAST MIMO 
systems with $n_t=n_r=8,16,24,32,64$ on flat fading ($L=K=1$). 
\# BP iterations = 20, $\alpha_m=0.4$.}
\label{fig11}
\vspace{-4mm}
\end{figure}

\begin{figure}
\hspace{-3mm}
\includegraphics[width=3.75in, height=2.90in]{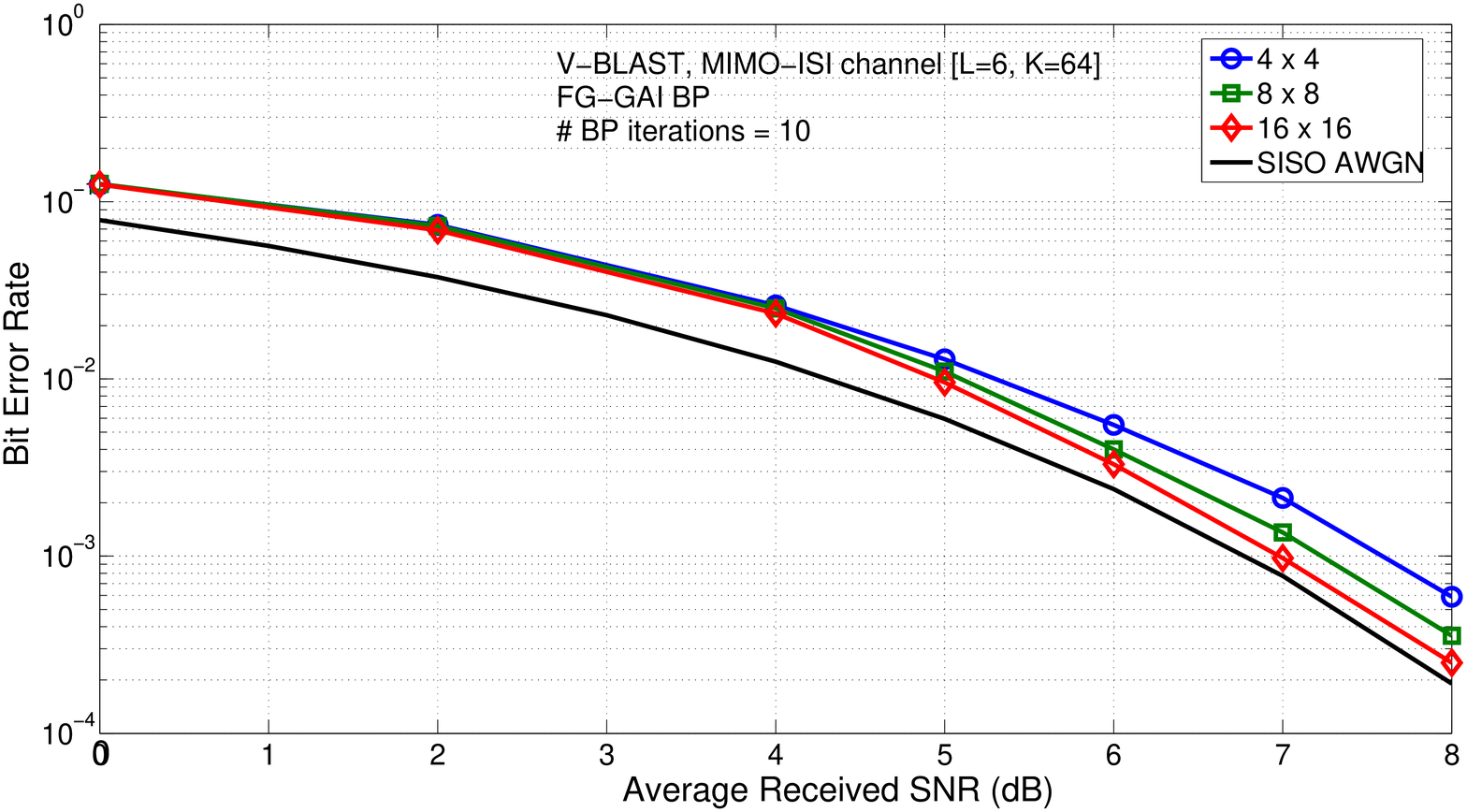}
\vspace{-6mm}
\caption{BER performance of the FG-GAI BP algorithm in MIMO-ISI channels 
with for $[L=6,K=64]$ for $n_t=n_r=4,8,16$. Uniform power delay
profile, \# BP iterations = 10, $\alpha_m=0.4$.}
\label{fig12}
\end{figure}

\begin{figure}
\hspace{-3mm}
\includegraphics[width=3.75in, height=2.90in]{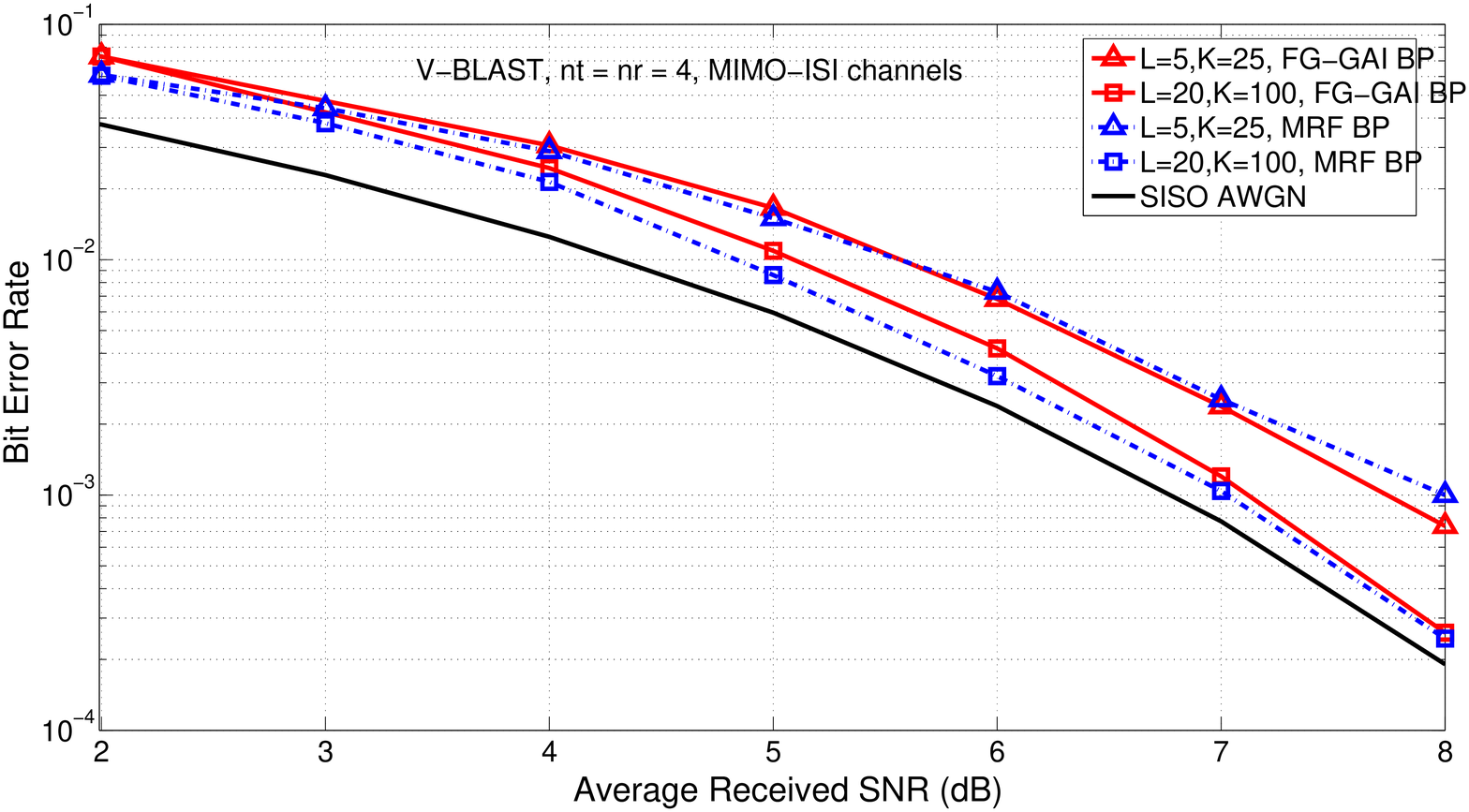}
\vspace{-6mm}
\caption{Comparison of the BER performances of the MRF BP and FG-GAI
BP algorithms in MIMO-ISI channels with $n_t=n_r=4$, $[L=5,K=25]$,
$[L=20,K=100]$, uniform power delay profile. }
\label{fig13}
\end{figure}

Figure \ref{fig11} shows the simulated BER performance of the FG-GAI BP
algorithm
in $n_t\times n_r$ V-BLAST MIMO with $n_t=n_r=8,16,24,32,64$ and BPSK on 
flat fading ($L=K=1$). The number of BP iterations and and message 
damping factor used are 10 and 0.4, respectively. We observe that, like 
the MRF approach, the FG-GAI approach also exhibits 
large-dimension behavior; e.g., $32\times 32$ and $64\times 64$ V-BLAST
systems perform close to unfaded SISO AWGN performance. Similar 
large-dimension behavior is shown in Fig. \ref{fig12} in MIMO-ISI channels 
with $L=6$ and $K=64$ for $n_t=n_r=4,8,16$; i.e., BERs move increasingly 
closer to unfaded SISO AWGN BER for increasing $Kn_t= 256,512,1024$.
Figure \ref{fig13} presents a comparison of the performances achieved
by the MRF and FG-GAI approaches for the following
system settings: $n_t=n_r=4$, $[L=5,K=25]$, $[L=20,K=100]$, and BPSK.
It can be seen that, for these system settings, the FG with GAI approach
performs almost the same as the MRF approach, at one order
lesser complexity than that of the MRF approach.

\begin{figure}
\hspace{-3mm}
\includegraphics[width=3.75in, height=2.90in]{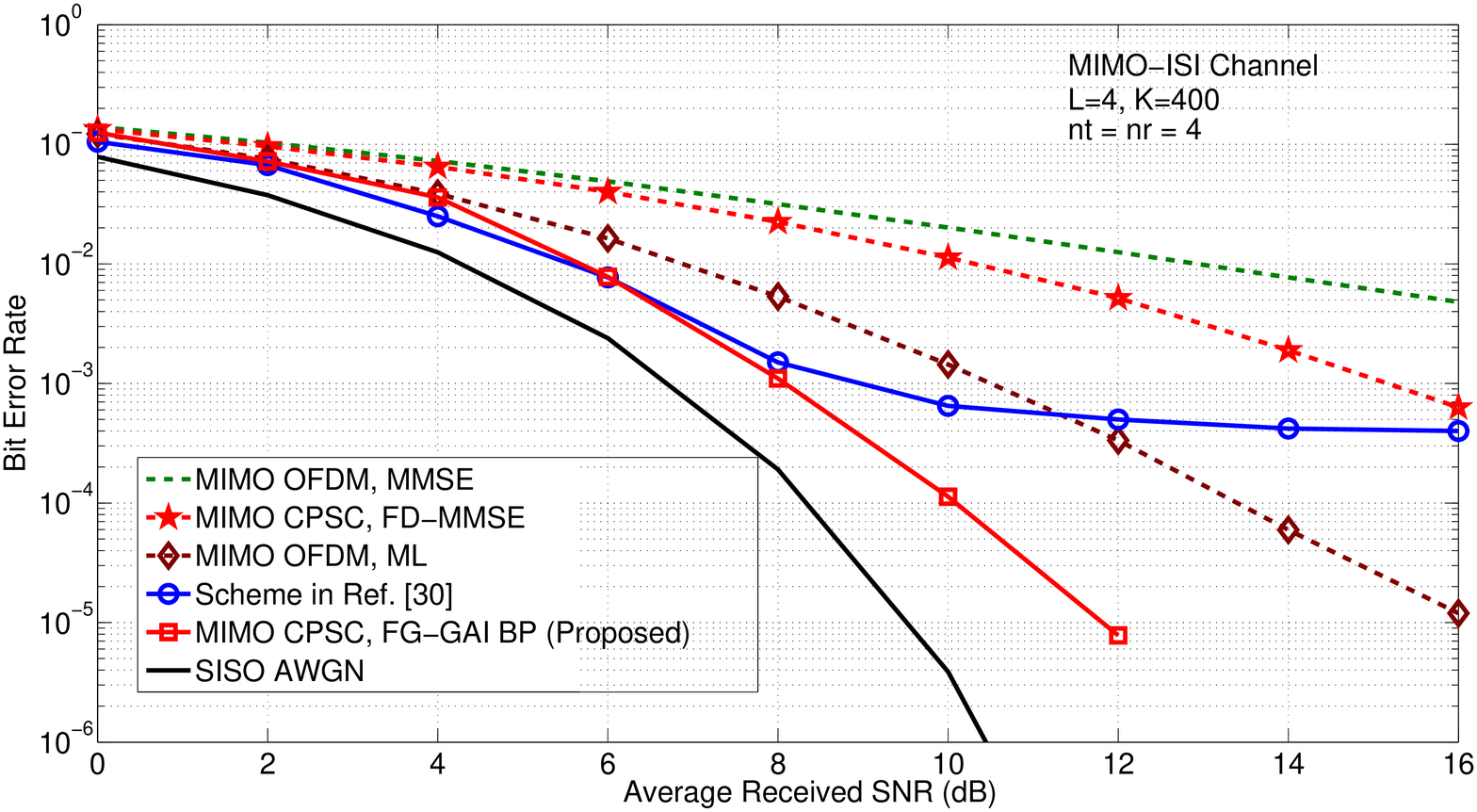}
\vspace{-6mm}
\caption{Comparison of the BER performances of the FG-GAI BP 
scheme and the scheme in \cite{wo} in MIMO-ISI channels with $n_t=n_r=4$, 
$[L=4,K=400]$, uniform power delay profile. }
\label{fig13a}
\end{figure}

Figure \ref{fig13a} presents a comparison of the performances achieved
by the proposed scheme and the scheme in \cite{wo} for $n_t=n_r=4$, 
$[L=4,K=400]$, and BPSK. It can be seen that while the scheme in 
\cite{wo} exhibits an error floor, the proposed scheme avoids flooring 
and achieves much better performance. Such good performance is achieved 
because equalization is done jointly on all the $Kn_t$ symbols in a 
frame. The complexity of the scheme in \cite{wo} is $O(Ln_t)$, whereas 
the complexity of the proposed scheme is $O(Kn_t)$. Though $K>L$, the 
linear complexity of the proposed scheme in $K$ is still very attractive. 
Also, as with MRF BP, the FG-GAI BP algorithm in MIMO-CPSC performs 
significantly better than MIMO-OFDM even with ML detection.

\section{Hybrid Algorithms Using BP and Local Neighborhood Search for $M$-QAM}
\label{sec5}
The BP algorithms proposed in the previous two sections are for BPSK 
modulation, i.e., for ${\bf x} \in \{\pm 1\}^{Kn_t}$. They can work for 
4-QAM also by viewing the transmit symbol vector to be in $\{\pm 1\}^{2Kn_t}$. 
Low-complexity algorithms for detection/equalization for higher-order $M$-QAM, 
$M>4$, over large dimension MIMO-ISI channels are of interest. A BP based 
algorithm that is suited for higher-order QAM in MIMO has been reported 
recently in \cite{gta}. The algorithm in \cite{gta} uses a Gaussian tree 
approximation (GTA) to convert the fully-connected graph representing the 
MIMO system into a tree, and carries out BP on the resultant approximate 
tree. We refer to this algorithm in \cite{gta} as the GTA BP algorithm. In 
this section, we take an alternate hybrid approach for efficient detection 
of $M$-QAM signals, where the proposed FG-GAI BP algorithm for BPSK is used 
to improve the $M$-QAM detection performance of local neighborhood search 
algorithms. Simulation results (Fig. \ref{fig16}) show that the proposed 
hybrid approach performs better than the GTA BP approach in \cite{gta}.

{\em Local Neighborhood Search Based Detection:}
Low complexity search algorithms that attempt to minimize the 
maximum-likelihood (ML) cost $\|{\bf r}-{\bf Hx}\|^2$, by limiting the 
search space to local neighborhood have been proposed for detection of 
$M$-QAM signals in MIMO -- e.g., tabu search (TS) algorithm 
\cite{tabu1}-\cite{isi_gcom09}. Such local neighborhood search algorithms 
have the advantage of low-complexity (e.g., TS algorithms, like the 
proposed MRF BP algorithm, has quadratic complexity in $Kn_t$), making 
them suited for large dimensions. However, their higher-order QAM 
performance is away from optimal performance. Here, we propose to improve 
the $M$-QAM performance of these search algorithms through the application 
of the proposed BP algorithms on the search algorithm outputs. This approach 
essentially improves the reliability of the output symbols from the local 
neighborhood search, thereby improving the overall BER performance. We
apply this hybrid approach to the reactive tabu search (RTS) algorithm
in \cite{isi_gcom09}.

{\em Hybrid RTS-BP Approach:}
In the 
following subsections, we first present a brief summary of the RTS algorithm
in \cite{isi_gcom09} and the motivation behind the proposed hybrid approach. 
Next, we present the proposed hybrid RTS-BP algorithm and its BER performance. 
Finally, we present a method to reduce complexity based on the knowledge of 
the simulated pdf of the RTS algorithm output.

\subsection{Reactive Tabu Search (RTS) Algorithm}
\label{sec_5a}
Here, we present a brief summary of the RTS algorithm in \cite{isi_gcom09}.
The RTS algorithm starts with an initial solution vector, defines a 
neighborhood around it (i.e., defines a set of neighboring vectors based 
on a neighborhood criteria), and moves to the best vector among the 
neighboring vectors (even if the best neighboring vector is worse, in 
terms of ML cost $\|{\bf r}-{\bf Hx}\|^2$, than the current solution 
vector); this allows the algorithm to escape from local minima. This 
process is continued for a certain number of iterations, after which the 
algorithm is terminated and the best among the solution vectors in all
the iterations is declared as the final solution vector. In defining the
neighborhood of the solution vector in a given iteration, the algorithm
attempts to avoid cycling by making the moves to solution vectors of the
past few iterations as `tabu' (i.e., prohibits these moves), which ensures
efficient search of the solution space. The number of these past iterations
is parametrized as the `tabu period,' which is dynamically changed depending
on the number of repetitions of the solution vectors that are observed in
the search path (e.g., increase the tabu period if more repetitions are
observed). The per-symbol complexity of the RTS algorithm is quadratic 
in $Kn_t$ for $n_t=n_r$.

\subsection{Motivation for Hybrid RTS-BP Algorithm}
\label{sec_5b}
The proposed hybrid RTS-BP approach is motivated by the following
two observations we made in our BER simulations of the RTS algorithm: 
$i)$ the RTS algorithm performed very close to optimum performance
in large dimensions for 4-QAM; however, its higher-order QAM performance
is far from optimal, and $ii)$ at moderate to high SNRs, when an RTS 
output vector is in error, the least significant bits (LSB) of the data 
symbols are more likely to be in error than other bits. An analytical 
reasoning for the second observation can be given as follows.

Let the transmitted symbols take values from $M$-QAM alphabet ${\mathbb A}$, 
so that ${\bf x} \in {\mathbb A}^{n_t}$ is the transmitted vector. 
Consider the real-valued system model corresponding to 
(\ref{eqn1}), given by 
${\bf r}' = {\bf H}' \, {\bf x}' + {\bf v}'$, where
\begin{eqnarray}
\label{SystemModelRealDef} \nonumber
\hspace{14mm}
{\bf H}' = \left[\begin{array}{cc}\Re({\bf H}) \hspace{2mm}-\Im({\bf H}) \\
\Im({\bf H})  \hspace{5mm} \Re({\bf H}) \end{array}\right],
\quad
{\bf r}' = \left[\begin{array}{c} \Re({\bf r}) \\ \Im({\bf r})
\end{array}\right],
\end{eqnarray}
\begin{eqnarray}
{\bf x}' = \left[\begin{array}{c} \Re({\bf x}) \\ \Im({\bf x})
\end{array}\right],
\quad
{\bf v}' = \left[\begin{array}{c} \Re({\bf v}) \\ \Im({\bf v})
\end{array}\right].
\end{eqnarray}
${\bf x}'$ is a $2Kn_t\times 1$ vector; 
$[x_1',\cdots,x_{Kn_t}']$ can be viewed to be from an underlying 
$M$-PAM signal set, and so is $[x_{Kn_t+1}',\cdots,x_{2Kn_t}']$.
Let
${\mathbb B} = \{a_1,a_2,\cdots,a_M\}$ denote the $M$-PAM alphabet
that $x_i'$ takes its value from. 

Let $\widehat{{\bf x}}'$ denote the detected output vector from the 
RTS algorithm corresponding to the transmitted vector ${\bf x}'$.
Consider the expansion of the $M$-PAM symbols in terms of $\pm 1$'s, 
where we can write the value of each entry of $\widehat{{\bf x}}'$ as
a linear combination of $\pm 1$'s as
\begin{eqnarray}
\label{linearComb}
\widehat{x}_i'&=&\sum_{j=0}^{N-1} 2^j \, \widehat{b}_i^{(j)}, \,\,\,\,\,\, i=1,\cdots,2Kn_t,
\end{eqnarray}
where $N=\log_2 M$ and $\widehat{b}_i^{(j)} \in \{\pm 1\}$. We note that
the RTS algorithm outputs a local minima as the solution vector. So,
$\widehat{\bf{x}}'$, being a local minima, satisfies the following conditions:

\vspace{-2mm}
{\small
\begin{eqnarray}
\Vert {\bf r}'-{\bf H}'\widehat{\bf x}'\Vert^{2} \, \le \, \Vert{\bf r}'-{\bf H}'(\widehat{\bf x}'+\lambda_i\textbf{e}_i)\Vert^{2}, \,\, \forall i=1,\cdots,2Kn_t,
\label{localminima}
\end{eqnarray}
}

\vspace{-4mm}
\hspace{-5mm}
where $\lambda_i = (a_q - \widehat{x}_i'), \, q=1,\cdots,M$,
and $\textbf{e}_i$ denotes the $i$th column of the identity matrix.
Defining ${\bf F}' \Define {\bf H}'^T{\bf H}'$
and denoting the $i$th
column of ${\bf H}'$ as ${\bf h}_i$,  the conditions in
(\ref{localminima}) reduce to
\begin{eqnarray}
2\lambda_i{\bf r}'^{T}{\bf h}_i & \le & 2\lambda_i({\bf H}'\widehat{\bf x}')^T {\bf h}_i + \lambda_i^2 f_{ii},
\label{reducedcondition}
\end{eqnarray}
where $f_{ij}$ denotes the $(i,j)$th element of ${\bf F}'$.
Under moderate to high SNR conditions, ignoring the noise,
(\ref{reducedcondition}) can be further reduced to
\begin{eqnarray}
2({\bf x'}-\widehat{\bf x}')^T {\bf f}_i \, \mbox{sgn}(\lambda_i) & \le & \lambda_i f_{ii} \, \mbox{sgn}(\lambda_i),
\label{finalcondn1}
\end{eqnarray}
where ${\bf f}_i$ denotes the $i$th column of ${\bf F}'$.
For Rayleigh fading, $f_{ii}$ is chi-square distributed with $2KN_t$
degrees of freedom with mean $KN_t$. Approximating the distribution of
$f_{ij}$ to be normal with mean zero and variance $\frac{KN_t}{4}$ for
$i\ne j$
by central limit theorem, we can drop the $\mbox{sgn}(\lambda_i)$ in
(\ref{finalcondn1}). Using the fact that the minimum value of
$|\lambda_i|$ is 2, (\ref{finalcondn1}) can be simplified as
\vspace{-1mm}
\begin{eqnarray}
\sum_{x_j'\ne \widehat{x}_{j}'} \Delta_j f_{ij} & \le & f_{ii},
\label{finalcondn2}
\end{eqnarray}
where $\Delta_j=x_j'-\widehat{x}_j'$.
Also, if $x_i'=\widehat{x}_i'$, by the normal approximation in the above
\begin{eqnarray}
\sum_{x_j'\ne \widehat{x}_{j}'} \Delta_j f_{ij} & \sim &
{\mathcal N}\Big(0,\frac{KN_t}{4}{\sum_{{x_j'} \ne \widehat{x}_{j}'}}\Delta_j^2\Big).
\label{finalline}
\end{eqnarray}
Now, the LHS in (\ref{finalcondn2}) being normal with variance proportional
to $\Delta_j^2$ and the RHS being positive, it can be seen that $\Delta_i$,
$\forall i$ take smaller values with higher probability. Hence, the symbols
of $\widehat{\bf x}'$ are nearest Euclidean neighbors of their corresponding
symbols of the transmitted vector with high probability\footnote{Because
$x_i'$'s and $\widehat{x}_i'$'s take values from $M$-PAM alphabet,
$\widehat{x}_i'$ is said to be the Euclidean nearest neighbor of
$x_i$ if $|x_i'-\widehat{x}_i'| = 2$.}.
Now, because of the symbol-to-bit mapping in (\ref{linearComb}),
$\widehat{x}_i'$ will differ from its nearest Euclidean neighbors
certainly in the LSB position, and may or may not differ in other
bit positions. Consequently,
the LSBs of the symbols in the RTS output $\widehat{{\bf x}}'$ are
least reliable.

The above observation then led us to consider improving the reliability
of the LSBs of the RTS output using the proposed FG-GAI BP algorithm 
presented in Section \ref{sec4}, and iterate between RTS and FG-GAI BP 
as follows.

\begin{figure}
\centering
\includegraphics[width=3.45in, height=1.2in]{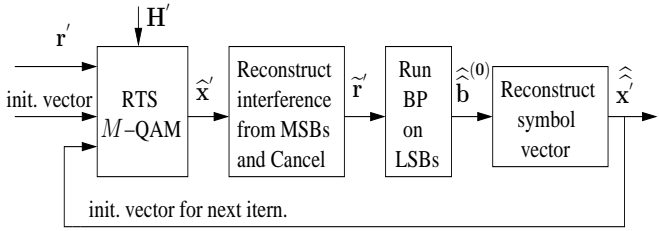}
\vspace{-4mm}
\caption{Hybrid RTS-BP algorithm.}
\label{fig14}
\end{figure}

\subsection{Proposed Hybrid RTS-BP Algorithm}
\label{sec_5c}
Figure \ref{fig14} shows the block schematic of the proposed hybrid
RTS-BP algorithm. The following four steps constitute the proposed
algorithm.
\begin{itemize}
\item {\em Step 1:}
Obtain $\widehat{{\bf x}}'$ using the RTS algorithm. Obtain the
output bits $\widehat{b}_i^{(j)}$, $i=1,\cdots,2Kn_t$,
$j=0,\cdots,N-1$, from $\widehat{{\bf x}}'$ and (\ref{linearComb}).
\item {\em Step 2:}
Using the $\widehat{b}_i^{(j)}$'s from Step 1, reconstruct the
interference from all bits other than the LSBs \big(i.e., interference
from all bits other than $\widehat{b}_i^{(0)}$'s\big) as
\begin{eqnarray}
\widetilde{{\bf I}} & = & \sum_{j=1}^{N-1} 2^{j}\, {\bf H}' \, \widehat{{\bf b}}^{(j)},
\label{intf}
\end{eqnarray}
where $\widehat{{\bf b}}^{(j)} = \big[\widehat{b}_1^{(j)}, \widehat{b}_2^{(j)}, \ldots, \widehat{b}_{2Kn_t}^{(j)} \big]^T$. Cancel the reconstructed
interference in (\ref{intf}) from {\bf r} as
\begin{eqnarray}
\widetilde{{\bf r}}' & = & {\bf r}' - \widetilde{{\bf I}}.
\end{eqnarray}
\item {\em Step 3:}
Run the FG-GAI BP algorithm in Section \ref{sec4} on the vector
$\widetilde{{\bf r}}'$ in Step 2, and obtain an estimate of the
LSBs. Denote this LSB output vector from FG-GAI BP as
$\widehat{\widehat{\bf b}}^{(0)}$.
Now, using $\widehat{\widehat{\bf b}}^{(0)}$ from the BP output,
and the $\widehat{\bf b}^{(j)}$, $j=1,\cdots,N-1$ from the RTS
output in Step 1, reconstruct the symbol vector as
\begin{eqnarray}
\widehat{\widehat{{\bf x}'}} & = &  \widehat{\widehat{\bf b}}^{(0)} \,+ \sum_{j=1}^{N-1} 2^{j}\, \, \widehat{{\bf b}}^{(j)}.
\end{eqnarray}
\item {\em Step 4:}
Repeat Steps 1 to 3 using $\widehat{\widehat{{\bf x}'}}$
as the initial vector to the RTS algorithm.
\end{itemize}
The algorithm is stopped after a certain number of iterations between
RTS and BP. Our simulations showed that two iterations between RTS and
BP are adequate to achieve good improvement; more than two iterations
resulted in only marginal improvement for the system parameters
considered in the simulations. Since the complexity of BP part of
RTS-BP is less than that of the RTS part, the order of complexity of
RTS-BP is same as that of RTS, $O(K^2n_t^2)$.

\begin{figure}
\hspace{-3mm}
\includegraphics[width=3.75in, height=2.90in]{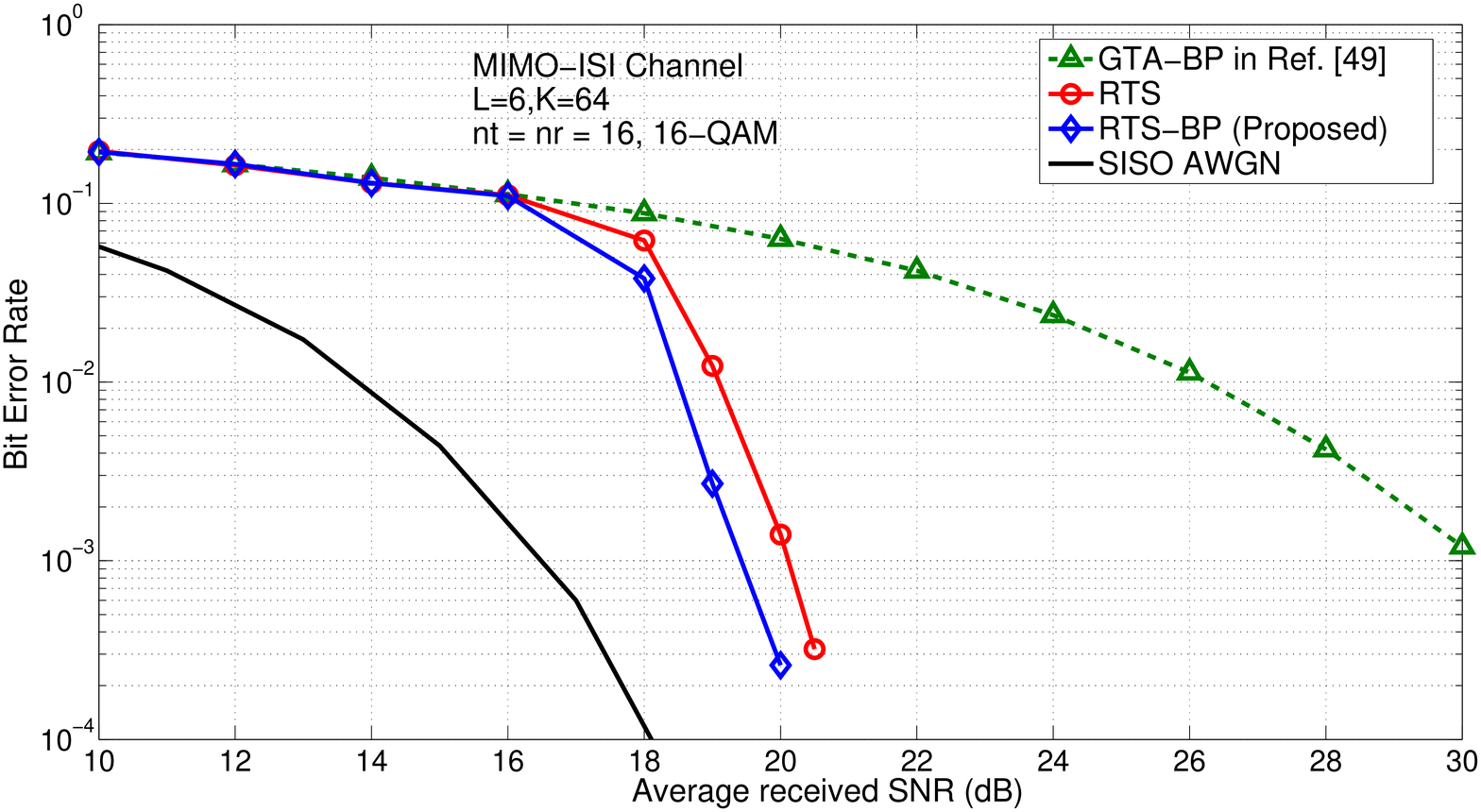}
\vspace{-6mm}
\caption{BER performance comparison between the RTS-BP (proposed), 
RTS, and GTA-BP (in \cite{gta}) in $16\times 16$ V-BLAST MIMO with 16-QAM 
in MIMO-ISI channel with $L=6$, $K=64$, uniform power-delay profile.}
\label{fig15}
\end{figure}

\subsection{Simulation Results}
\label{sec_5d}
Figure \ref{fig15} shows the BER performance of the proposed hybrid RTS-BP 
algorithm in comparison with those of the RTS algorithm and the GTA-BP 
algorithm in \cite{gta} in $16\times 16$ V-BLAST MIMO with 16-QAM on a 
frequency selective channel with $L=6$ equal energy multipath components 
and $K=64$ data vectors per frame. Because of the improvement of the 
reliability of LSBs due to BP run on them, the RTS-BP algorithm achieves 
better performance compared to RTS algorithm without BP. Also, both RTS-BP 
and RTS algorithms perform better than the GTA-BP in \cite{gta}.

\subsection{Complexity Reduction Using Selective BP}
\label{sec_5e}
\begin{figure}
\hspace{-3mm}
\includegraphics[width=3.75in, height=2.90in]{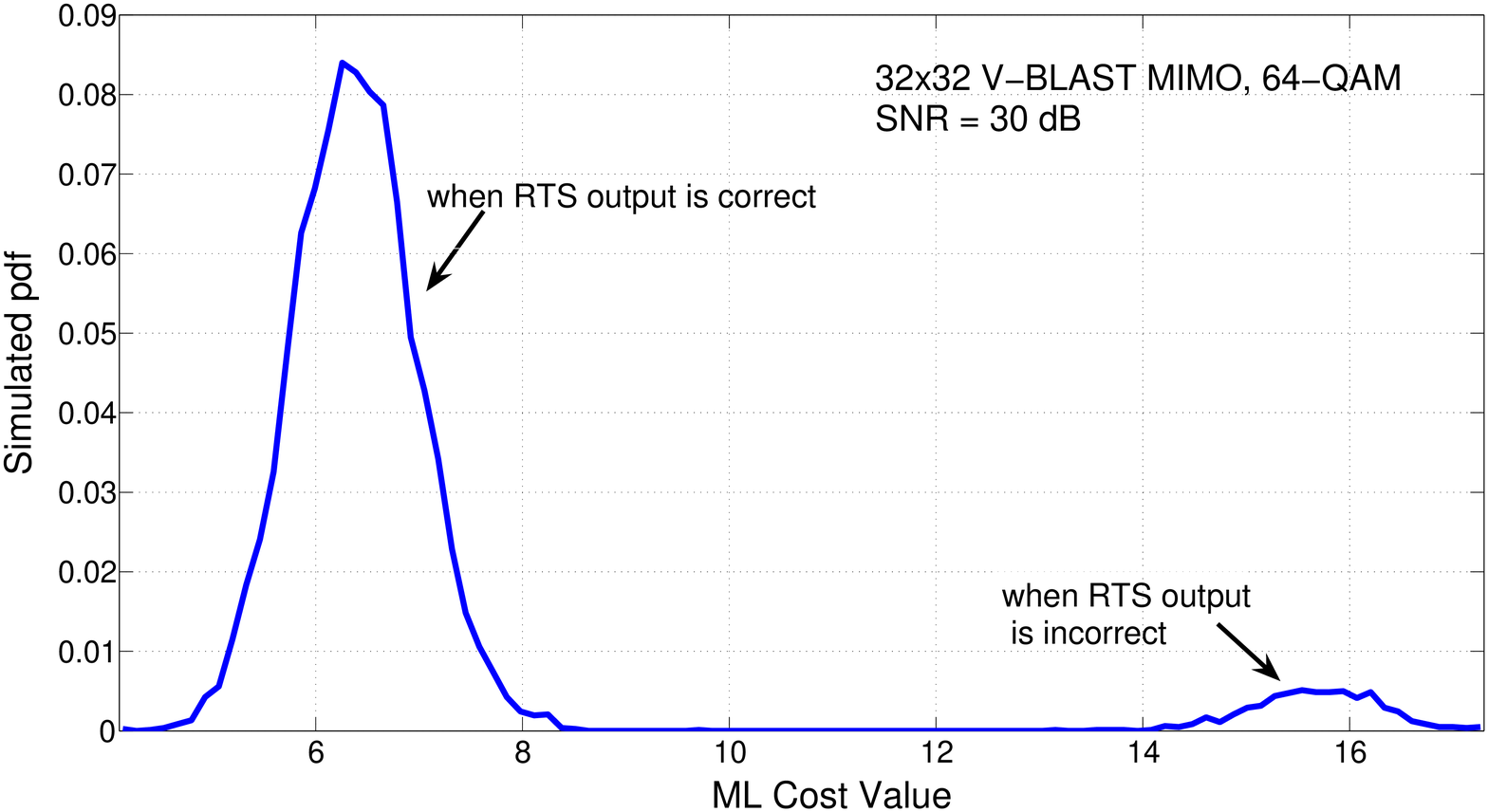}
\vspace{-6mm}
\caption{Simulated pdfs of $M_1$, the ML cost of the RTS output vector,
in a $32\times 32$ V-BLAST MIMO system with 64-QAM and SNR = 30 dB on
flat fading ($L=K=1$).
}
\label{fig16}
\end{figure}

In the proposed RTS-BP algorithm, the use of BP at the
RTS output was done unconditionally. Whereas the use of BP can improve
performance only when the RTS output is erroneous. So, the additional
complexity due to BP can be avoided if BP is not carried out whenever
the RTS output is error-free. To decide whether to use BP or not,
we can use the knowledge of the simulated pdf of the ML cost of the RTS
output vector, i.e., the pdf of
$M_1\Define \Vert{\bf r}'-{\bf H}'\widehat{{\bf x}}'\Vert$.
Figure \ref{fig16} shows the simulated pdf of $M_1$ for a $32\times 32$
V-BLAST MIMO system with 64-QAM at an SNR of 30 dB on flat fading ($L=K=1$). 
From Fig. \ref{fig16}, it is seen that a comparison of the value of $M_1$ 
with a suitable threshold can give an indication of the reliability of the 
RTS output. For example, the output is more likely to be erroneous if $M_1 >12$ 
in Fig. \ref{fig16}.

\begin{figure}
\hspace{-3mm}
\includegraphics[width=3.75in, height=2.90in]{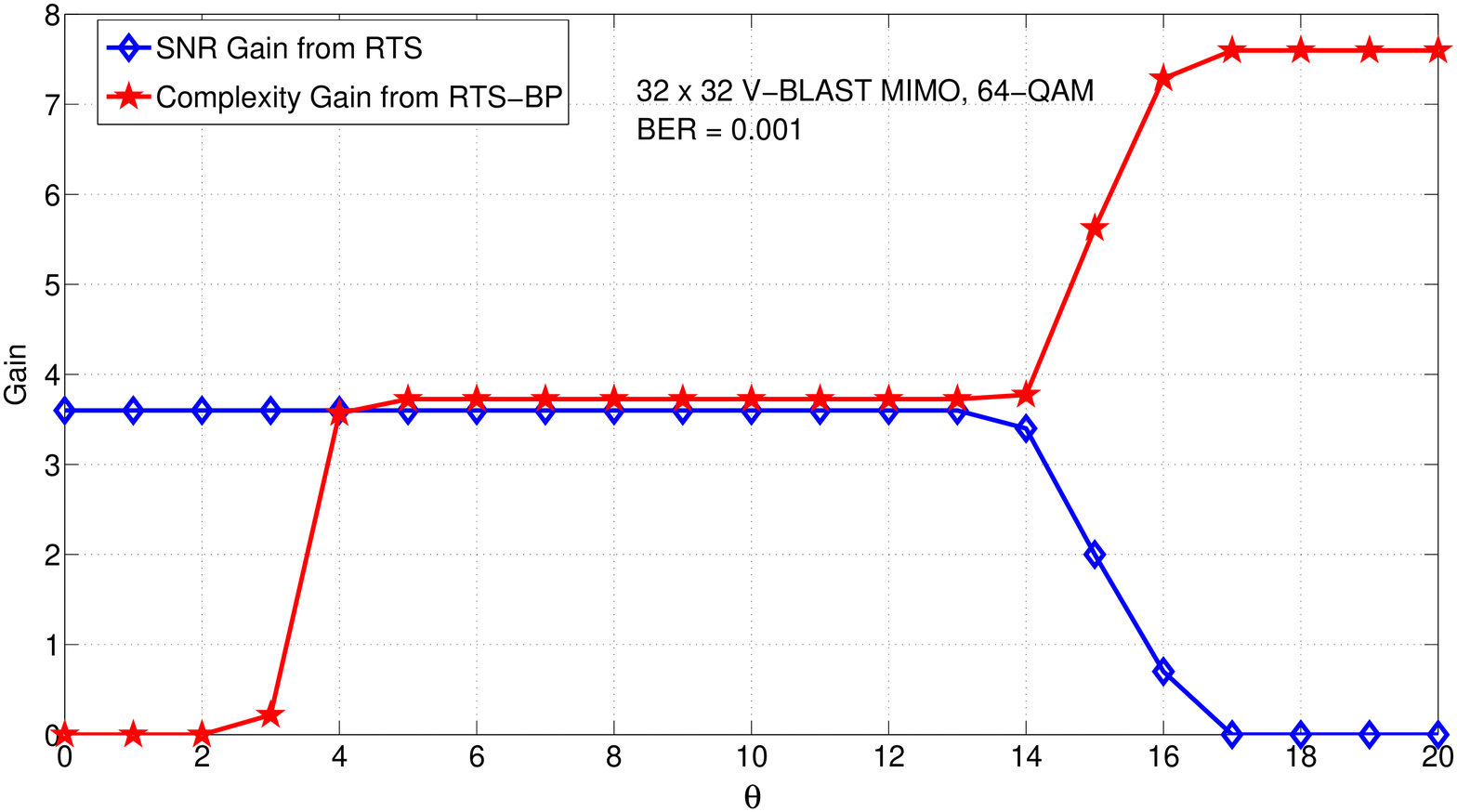}
\vspace{-6mm}
\caption{
SNR gain versus complexity gain trade-off in selectively using BP
as a function of $\theta$ in a $32\times 32$ V-BLAST MIMO system 
with 64-QAM at a BER of 0.001 on flat fading ($L=K=1$).
}
\label{fig17}
\end{figure}

Based on the above observation, we modify the RTS-BP algorithm as follows.
If $M_1 > \theta$, only then
BP algorithm is used; otherwise, the RTS output is taken as the final
output. The threshold $\theta$ has to be carefully chosen to achieve
good performance. It is seen that $\theta=0$ corresponds to the case of
unconditional RTS-BP, and $\theta=\infty$ corresponds to the case of RTS
without BP. For $\theta=\infty$, there is no additional complexity due
to BP, but there is no performance gain compared to RTS. For $\theta=0$,
performance gain is possible compared to RTS, but BP complexity
will be there for all realizations. So there exits a performance-complexity
trade off as a function of $\theta$. We illustrate this trade-off in
Fig. \ref{fig17} for a $32\times 32$ V-BLAST system with 64-QAM in
flat fading. For
this purpose, we define `SNR gain' in dB for a given threshold $\theta$
as the improvement in SNR achieved by RTS with selective BP using threshold
$\theta$ to achieve an uncoded BER of $10^{-3}$ compared to RTS without BP.
Likewise, we define `complexity gain' for a given $\theta$ as
$10\log_{10}(\beta)$, where $\beta$ is the ratio of the average number of
computations required to achieve $10^{-3}$ uncoded BER in unconditional
RTS-BP and that in RTS with selective BP using threshold $\theta$. In
Fig. \ref{fig17}, we plot these two gains on the y-axis as
a function of the threshold $\theta$. From this figure, we can observe that
for $\theta$ values less than 4, there is not much complexity gain since such
small threshold values invoke BP more often (i.e., the system behaves
more like unconditional RTS-BP). Similarly, for $\theta$ values
greater than 14, the system behaves more like RTS without BP; i.e., the
complexity gain is maximum but there is no SNR gain. Interestingly,
for $\theta$ values in the range 4 to 14, maximum SNR gain is retained
while achieving significant complexity gain as well.

\section{Conclusions}
\label{sec6}
In this paper, we demonstrated that belief propagation on graphical models 
including Markov random fields and factor graphs can be efficiently used to 
achieve near-optimal detection in large-dimension MIMO-ISI channels at 
quadratic and linear complexities in $Kn_t$. 
It was shown through simulations that damping of messages/beliefs in the
MRF BP algorithm can significantly improve the BER performance and 
convergence behavior. The Gaussian approximation of interference
we adopted in the factor graph approach is novel, which offered the 
attractive linear complexity in number of dimensions while achieving 
near-optimal performance in large dimensions. In higher-order QAM,
iterations between a tabu search algorithm and the proposed FG-GAI BP 
algorithm was shown to improve the bit error performance of the basic 
tabu search algorithm. Although we have demonstrated the proposed algorithms
in uncoded systems, they can be extended to coded systems as well, using 
either turbo equalization or joint processing of the entire coded symbol 
frame based on low-complexity graphical models. Finally, a theoretical 
analysis of the convergence behavior and the bit error performance of 
the proposed BP algorithms is challenging, and remains to be studied.

\end{document}